\documentclass{article}
 
\usepackage{amssymb}
\usepackage{amscd}
\usepackage{bm}
\usepackage{graphics}
\usepackage{amsfonts}
\usepackage{amsthm}
\usepackage{amsmath}
\usepackage{latexsym}
\usepackage[all]{xy}
\usepackage{fancyhdr}
\usepackage{color}
\usepackage{cite}

\usepackage{latexsym,graphics,color}

\def\be{\begin{equation}}
\def\ee{\end{equation}}

\def\D{\mathcal D}
\def\dd{\mathbb D}
\def\gH{\mathfrak H}
\def\B{\mathcal B}
\def\P{\mathcal P}

\def\H{h}
\def\eq{\eqref}
\def\E{\mathcal E}
\def\F{\mathcal F}
\def\bc{\mathbb C}
\def\br{\mathbb R}
 
\def\ma{ \reflectbox{\it a}}
\def\na {{\reflectbox{\reflectbox{\it a}}}}
\def\aq{{ \reflectbox{\it aq}}}
\def\qa {{\reflectbox{\reflectbox{\it aq}}}}

\def\T{ \reflectbox{\it T}}
\def\K{\mathcal K}
\def\J{\mathcal J}
\def\L{\mathcal L}

\def\ad{{\rm ad}}
\def\dk {\partial_\kappa}
\def\d{\partial}
\def\jp{\frac{1}{2}}

\def\ups{\Upsilon}
\def\al{\alpha}

\def\1{{\mbox{\boldmath $1$}}}          %

\def\la{\langle}
\def\ra{\rangle}
\def\e{\epsilon}                        %
\def\jp{\frac{1}{2}}                    %
\def\al{\alpha}                         %

\definecolor{spec}{rgb}{0.0, 0.26, 0.15}

\begin{document}

\begin{flushright}
{}~
  
\end{flushright}

\vspace{1cm}
\begin{center}
{\large \bf Affine Poisson and affine quasi-Poisson T-duality}

\vspace{1cm}

{\small
{\bf Ctirad Klim\v{c}\'{\i}k}
\\
Aix Marseille Universit\'e, CNRS, Centrale Marseille\\ I2M, UMR 7373\\ 13453 Marseille, France}
\end{center}

\vspace{0.5 cm}
\centerline{\bf Abstract}
\vspace{0.5 cm}
\noindent  We generalize the Poisson-Lie T-duality by making use of the  structure of the affine Poisson group which is the concept introduced some time ago in Poisson geometry as a generalization of the Poisson-Lie group. We also  introduce a new notion of an affine quasi-Poisson group and show that it gives rise to a still more general T-duality framework.
We  establish for a  class of examples that this new T-duality is compatible with the renormalization group flow.

\vspace{2pc}

\section{Introduction}
\setcounter{equation}{0}

The Poisson-Lie T-duality \cite{KS95}  is the framework  which permits to construct  examples of dynamically equivalent non-linear $\sigma$-models living on geometrically non-equivalent backgrounds. The basic structural ingredient underlying this kind of  T-duality is the so called Drinfeld double $D$, which is a Lie group equipped with a bi-invariant metric of maximally Lorentzian (or split) signature and having two half-dimensional isotropic subgroups $K$ and $\tilde K$.  

The structure of $D$   induces  certain  Poisson brackets on each subgroup $K$ and $\tilde K$.  Those brackets are
called the Poisson-Lie ones and, remarkably, the Poisson-Lie bivectors which correspond to them  appear explicitely in the actions of the mutually  T-dual $\sigma$-models\footnote{The formulae \eqref{mmm} and \eqref{mmm'} are valid for the so called {\it perfect} Drinfeld doubles
for which there exists a diffeomorphism $\Upsilon: D\to K\times \tilde K$ composing
to identity with the group multiplication map $m:  K\times \tilde K\to D$, i.e. it must hold that $m\circ \Upsilon$ is the identity map from $D$ to $D$. In the present paper, we shall consider only the doubles $D$ verifying this property.}:
\be S=\jp\int d\tau\oint d\sigma     \biggl( \Bigl(E+\Pi(k)\Bigr)^{-1}\partial_+k k^{-1}, \partial_- k k^{-1}\biggr)_\D, \qquad k(\tau,\sigma)\in K;\label{mmm}\ee 
\be \tilde S=\jp\int d\tau\oint d\sigma    \biggl( \Bigl(\tilde E+\tilde \Pi(\tilde k)\Bigr)^{-1}\partial_+\tilde k \tilde k^{-1}, \partial_- \tilde k \tilde k^{-1}\biggr)_\D, \qquad \tilde k(\tau,\sigma)\in \tilde K.\label{mmm'}\ee 
Note that the respective targets of the $\sigma$-models \eqref{mmm} and \eqref{mmm'} are the subgroups $K$ and $\tilde K$ of the double $D$, the bilinear form $(.,.)_\D$ on the Lie algebra $\D$ is the split  metric evaluated at the unit element of the the group $D$, the linear operators $E:\tilde\K\to\K$ and $\tilde E:\K\to\tilde\K$ are constant, but the operators $\Pi(k):\tilde\K\to\K$ and $\tilde \Pi(\tilde k):\K\to\tilde\K$ depend on the points of the targets. Actually, the operators  $\Pi(k)$ and $\tilde \Pi(\tilde k)$ encode the Poisson-Lie brackets on $K$ and on $\tilde K$ by the formulae 
\be \{f_1,f_2\}_{K}(k)=(\nabla^l f_1,\Pi(k)\nabla^l f_2)_\D; 
\qquad \{\tilde f_1,\tilde f_2\}_{\tilde K}(\tilde k)=\left(\tilde\nabla^l \tilde f_1,\tilde\Pi(\tilde k)\tilde\nabla^l \tilde f_2\right)_\D.\label{171}\ee
Here $f_1,f_2$ and $\tilde f_1, \tilde f_2$ are, respectively, functions on the groups $K$ and $\tilde K$ and  the precise definitions of the $\tilde K$-valued  and the $\K$-valued 
right-invariant differential operators $\tilde\nabla$ and $\nabla$ are given in Eqs. \eqref{431} and \eqref{39}.
It was established in \cite{KS95,KS96a} that the T-duality relates the models \eqref{mmm} and \eqref{mmm'} if $\tilde E$ is inverse of $E$.  

The   duality existing in the realm of the Poisson-Lie groups expresses the fact that starting from a group $K$ and the Poisson-Lie structure $\Pi(k)$ on it, one can construct the dual group $\tilde K$ and the dual Poisson-Lie structure $\tilde\Pi(\tilde k)$ and repeating the same procedure with the pair $(\tilde K,\tilde \Pi(\tilde k))$ one gets back to the pair $(K,\Pi(k))$. It is truly remarkable that this purely geometric duality of the Poisson-Lie groups gets transported via the actions \eqref{mmm} and \eqref{mmm'} to the dynamical  T-duality in string theory. 

Apart from the Poisson-Lie duality, there exists another natural geometric duality in   Poisson geometry, the one  which  flips two Poisson-Lie groups associated to the so called the {\it affine Poisson group} $K$  \cite{DS,Lu,K07}. We speak then about geometric {\it affine Poisson duality} and  the purpose of the present paper is to convert it into a stringy  {\it affine Poisson T-duality}. The affine Poisson structure $\Pi^{\na}(k)$ on a group $K$  gives  rise naturally to the existence of three other Poisson manifolds : the so-called mirror affine Poisson 
group $(K,\Pi^{\ma}(k))$ and also  two (dual) Poisson-Lie groups $\left(\tilde K_L,\tilde\Pi^L(\tilde k_L)\right)$  and $\left(\tilde K_R,\tilde\Pi^R(\tilde k_R)\right)$. We  can construct four  $\sigma$-model actions  of the types \eqref{mmm}, \eqref{mmm'} for every of those four  Poisson structures 
\be  S_{\na}=\jp\int d\tau\oint d\sigma   \biggl(\Bigl(E^{\na}+\Pi^{\na}(k)\Bigr)^{-1}\partial_+k k^{-1}, \partial_- k k^{-1}\biggr)_{\D_R},\label{mm1}\ee
\be  S_{\ma}=\jp\int d\tau\oint d\sigma    \biggl( \Bigl(E^{\ma}+\Pi^{ \ma}(k)\Bigr)^{-1}\partial_+k k^{-1}, \partial_- k k^{-1}\biggr)_{\D_L},\label{mm2}\ee
 \be \tilde S_{\na}=\jp\int d\tau\oint d\sigma   \biggl(\Bigl(\tilde E^\na+\tilde\Pi^R(\tilde k_R)\Bigr)^{-1}\partial_+\tilde k_R\tilde k_R^{-1}, \partial_-\tilde k_R\tilde k_R^{-1}\biggr)_{\D_R}.\label{mm3}\ee  
  \be \tilde S_{\ma}=\jp\int d\tau\oint d\sigma   \biggl(\Bigl(\tilde E^{\ma}+\tilde\Pi^L(\tilde k_L)\Bigr)^{-1}\partial_+\tilde k_L\tilde k_L^{-1}, \partial_-\tilde k_L\tilde k_L^{-1}\biggr)_{\D_L}.\label{mm4}\ee
  
The first  result of this paper is the statement that for a large class of the affine Poisson groups
 $(K,\Pi^\na(k))$ one can choose the  linear operator $E^\na$ and, in terms of it, all other operators  $E^{\ma},\tilde E^\na$ and $\tilde E^{\ma}$ in such a way that all four  $\sigma$-models
 \eqref{mm1}, \eqref{mm2}, \eqref{mm3} and \eqref{mm4} 
 become pairwise T-dual to each other.  The most interesting T-duality is that
 relating the models \eqref{mm3} and \eqref{mm4}, since, as it will turn out,  this is the only one which cannot be reduced to the standard Poisson-Lie T-duality. However, if the affine Poisson structure $\Pi^\na(k)$ is equal to its mirror $\Pi^{\ma}(k)$,  then we  recover the standard Poisson-Lie T-duality, i.e.    
   the $\sigma$-models  \eqref{mm1}, \eqref{mm2} merge to give
the model \eqref{mmm} and the $\sigma$-models  \eqref{mm3}, \eqref{mm4} merge to give the model \eqref{mmm'}.

The second   result of this paper is the generalization of the pairwise T-duality of the four   $\sigma$-models  \eqref{mm1}, \eqref{mm2}, \eqref{mm3} and \eqref{mm4}  to the case where the structure $\Pi^{\na}(k)$ is not affine  Poisson but only affine quasi-Poisson. We stress that the  concept of the affine quasi-Poisson group is (apparently) a new one and we introduce it in the present paper. Why do we do  it? Because while proving the affine Poisson T-duality of the four  $\sigma$-models  \eqref{mm1}, \eqref{mm2}, \eqref{mm3} and \eqref{mm4},
we have discovered that the duality holds even if the structure $\Pi^\na(k)$ 
does not give rise to a bracket of fonctions  satisfying the Jacobi identity. We have then   worked out what kind of violation of the Jacobi identity is compatible with the T-duality 
and found that it is one which is well-known in the literature  
under the name of the quasi-Poisson geometry. Because it  is true, at the same time, that structures  which would be simultaneously affine and quasi-Poisson have not yet been  introduced, we devote one section of the present article  to the definition of the affine quasi-Poisson groups and to the study of their basic properties.  

The plan of the paper is as follows: In Section 2 and 3, we review the concept of T-duality as the symplectomorphism relating the phase spaces of two mutually dual $\sigma$-models,   we recall   the elements of the standard Poisson-Lie T-duality and we  collect some useful facts from the mathematical literature about  the theory of the affine Poisson groups. In Section 4,   we establish the T-duality of the four $\sigma$-models
\eqref{mm1}, \eqref{mm2}, \eqref{mm3} and  \eqref{mm4} when $K$ is the affine Poisson group. In Section 5, we introduce and study the concept of the affine quasi-Poisson group and, in Section 6, we establish the pairwise T-duality of the four  models \eqref{mm1}, \eqref{mm2}, \eqref{mm3} and  \eqref{mm4}
in  the case where $K$ is the affine quasi-Poisson group. In Section 7, we   expose the theory of dressing cosets \cite{KS96b} and show how the affine Poisson and the affine quasi-Poisson T-dualities can be interpreted from this vantage point. In Section 8, we show that for a particular class of the affine Poisson groups, the affine Poisson T-duality is compatible with the renormalisation group flow.  We finish by an outlook where  few open problems are formulated.

\section{$\E$-models, Poisson-Lie T-duality and outline of generalization} 
  \setcounter{equation}{0}

By definition, the Poisson-Lie T-duality is a symplectomorphism  that maps the phase space and the Hamiltonian of the   non-linear $\sigma$-model \eqref{mmm} onto the phase space and the Hamiltonian of the dual  $\sigma$-model \eqref{mmm'}. We shall  refer to those models as to the Poisson-Lie $\sigma$-models. 

Recall that the basic structural ingredients needed to construct the dual pair of the Poisson-Lie $\sigma$-models \eqref{mmm} and \eqref{mmm'} are  a $2d$-dimensional Lie algebra $\D$ (the Drinfeld double Lie algebra),  a symmetric ad-invariant non-degenerate bilinear form $(.,.)_\D$ on $\D$ with the split signature $(d,d)$, two $d$-dimensional Lie subalgebras $\K$ and $\tilde\K$ of $\D$ such that the restriction of the form $(.,.)_\D$  vanish on both of them, and, finally, a $d$-dimensional linear subspace $\E$ of $\D$ such that the restriction of $(.,.)_\D$ on  $\E$ is strictly positive definite.  All those data are needed in order to write down a duality invariant description of the common first order dynamics of the Poisson-Lie $\sigma$-models \eqref{mmm} and \eqref{mmm'} in terms of the so called 
$\E$-models \cite{KS96a, K15}.

The points of the phase space  of the $\E$-model are maps $l:S^1\to D$ from a circle parametrized by the variable $\sigma$ into the Drinfeld double $D$, or, in other words, they are the elements of the loop group $LD$ of the Drinfeld double. The symplectic form $\omega_{LD}$ of the $\E$-model is given by the following simple formula
\be \omega_{LD}:=-\jp \oint d\sigma  (l^{-1}dl,\partial_\sigma(l^{-1}dl))_\D,\label{342}\ee
where the symbol $\oint$ stands for integration over the loop parameter. 
Finally, the Hamiltonian $Q_\E$ is given by the formula 
\be Q_\E=\jp\oint d\sigma  (\partial_\sigma ll^{-1},\E \partial_\sigma ll^{-1})_\D.\label{345}\ee
Recall that we have defined $\E$ as the $d$-dimensional subspace of the $2d$-dimensional Drinfeld double Lie algebra $\D$, such that the restriction
of the bilinear form $(.,.)_\D$ on $\E$ is strictly positive definite. By the abuse of notation, we denote by $\E$ also the  linear operator 
$\E:\D\to\D$, self-adjoint with respect to the bilinear form
$(.,.)_\D$, which has the subspace $\E$ as the eigenspace for the eigenvalue $+1$ and the orthogonal complement subspace $\E^\perp$  as the eigenspace for the eigenvalue $-1$. 

For the purpose of this paper, it will be sufficient to consider a simpler variant of the Poisson-Lie T-duality (actually, the one originally introduced in \cite{KS95,KS96a})  for which the Drinfeld double is perfect in the sense of the Footnote 1. In this case, it is particularly simple to construct the symplectomorphisms corresponding to the arrows in the following chain of dynamical systems:
 \be (T^*LK,\omega,\H_\E)\longleftrightarrow (LD,\omega_{LD},Q_\E)\longleftrightarrow (T^*L\tilde K,\tilde\omega,\tilde\H_\E).\label{Sch1'}\ee
 Here $\omega$ is the canonical symplectic form on the cotangent bundle of the loop group 
 $T^*LK$:
 \be \omega =d\left(\oint d\sigma \la \beta, k^{-1}dk\ra\right),\qquad k\in LK, \quad \beta
 \in L\K^*\label{sf}\ee
 and the arrow pointing from $(LD,\omega_{LD},Q_\E)$ to the left is obtained by writing $l\in LD$
 as the product \be l=k\tilde h, \qquad k\in LK, \quad \tilde h\in L\tilde K.\label{dd}\ee 
 Indeed, inserting the decomposition \eqref{dd} into \eqref{342}, we obtain easily
 \be \omega_{LD}=d\left(\oint d\sigma(\partial_\sigma \tilde h\tilde h^{-1},k^{-1}dk)_\D\right).\label{sf1}\ee
 To make coincide the forms \eqref{sf} and \eqref{sf1}, it is now sufficient to remark that  $\partial_\sigma \tilde h\tilde h^{-1}$ is the element of $L\tilde\K$ and can be therefore identified with
 $\beta\in L\K^*$ via the non-degenerate bilinear form $(.,.)_\D$.

Inserting the decomposition  \eqref{dd}  into \eqref{345}, we obtain also the Hamiltonian $h_\E$ featuring in the scheme 
\eqref{Sch1'}:
\be h_\E(k,\beta) =\jp \oint d\sigma(\partial_\sigma kk^{-1}+ k\beta k^{-1},\E(\partial_\sigma kk^{-1}+ k\beta k^{-1}))_\D,\label{hpm}\ee
where we view $\beta$ as the element of $L\tilde\K$.

The first order action of the dynamical system $(T^*LK,\omega,\H_\E)$  is now given by the data
\eqref{sf} and \eqref{hpm}:
\be S_\E=\int d\tau \oint d\sigma   (\beta,k^{-1}\partial_\tau k)_\D- \int d\tau h_\E(k,\beta). \label{foa}\ee
The dependence of $S_\E$ on $\beta$ is quadratic, it is therefore easy to eliminate $\beta$  which gives the second order action  of the Poisson-Lie $\sigma$-model  \eqref{mmm}: 
 \be S_\E(k)=\jp\int d\tau \oint d\sigma  
 \left(\left(E+\Pi(k)\right)^{-1}\partial_+kk^{-1}, \partial_-kk^{-1}\right)_\D.\label{ea}\ee
Note that $E:\tilde\K\to\K$ is the linear operator such that its graph $\{\tilde x+E\tilde x,\tilde x\in\tilde\K\}$ coincides with the subspace $\E$ in the double $\D$ and we recall that $\Pi(k):\tilde\K\to\K$ encodes the Poisson-Lie bracket of two functions $f_1,f_2$  on the group $K$ in the sense of the formula:
\be \{f_1,f_2\}_{K}(k)=(\nabla^l f_1,\Pi(k)\nabla^l f_2)_\D.\label{spls'}\ee
Here $\nabla^{l}$ is $\tilde\K$-valued differential operator acting on the functions on $K$  as
\be (\nabla^l f, x)_\D(k):= (\nabla^l_{x}f)(k)\equiv \frac{df(e^{sx}k)}{ds}\bigg\vert_{s=0}, \qquad x\in\K.\label{431}\ee
 
The Poisson-Lie $\sigma$-model dual to the model \eqref{ea} can be obtained from \eqref{342} and \eqref{345}
in the same way starting from the dual  decomposition of the Drinfeld double:
\be l=\tilde kh, \qquad  \tilde k\in L\tilde K,\quad h\in LK,\label{ddd}\ee 
The result is the Poisson-Lie $\sigma$-model \eqref{mmm'}
\be \tilde S_\E(\tilde k)=\jp\int d\tau \oint d\sigma\biggl(\Bigl(\tilde E+\tilde\Pi(\tilde k)\Bigr)^{-1}\partial_+\tilde k \tilde k^{-1}, \partial_- \tilde k \tilde k^{-1}\biggr)_\D,\label{ead}\ee
Here $\tilde E:\K\to\tilde\K$ is the linear operator such that its graph $\{x +\tilde Ex, x\in\K\}$ coincides with the subspace $\E$ in the double $\D$, which in fact  means that $\tilde E=E^{-1}$.  Of course,  $\tilde\Pi(\tilde k):\K\to\tilde\K$ encodes  the Poisson-Lie structure on the group $\tilde K$ in the sense of the formula   
\be \{\tilde f_1,\tilde f_2\}_{\tilde K}(\tilde k)=\left(\tilde\nabla^l \tilde f_1,\tilde\Pi(\tilde k)\tilde\nabla^l \tilde f_2\right)_\D,\label{dspl}\ee
where $\tilde\nabla^{l}$ is $\K$-valued differential operator acting on the functions on $\tilde K$  as
\be (\tilde\nabla^l \tilde f,\tilde x)_\D(\tilde k):= (\tilde\nabla^l_{\tilde x}\tilde f)(\tilde k)= \frac{d\tilde f(e^{s\tilde x}\tilde k)}{ds}\bigg\vert_{s=0}, \qquad   \tilde x\in \tilde\K.\label{39}\ee

We now draw another scheme, which replaces that \eqref{Sch1'} and encapsulates   the generalization of the Poisson-Lie T-duality introduced in the present paper: 
 
\begin{align}   (T^*LK,\omega,\H_{\E_L})\longleftrightarrow (LD_L,&\omega_{LD_L},Q_{\E_L})\longleftrightarrow (T^*L\tilde K_L,\tilde\omega_L,\tilde\H_{\E_L})   \nonumber
\\ \updownarrow  \hskip 2.7cm &\updownarrow\hskip 9pc\updownarrow\label{Sch2} \\
(T^*LK,\omega,\H_{\E_R})\longleftrightarrow (LD_R,&\omega_{LD_R},Q_{\E_R})\longleftrightarrow (T^*L\tilde K_R,\tilde\omega_R,\tilde\H_{\E_R})  .\nonumber
\end{align}
 The horizontal and the vertical bidirectional arrows in the scheme \eqref{Sch2} stand for the Hamiltonian-preserving symplectomorphisms between the dynamical systems represented by the triples  $(P,\omega,\H)$, where $P$ is the phase space of the dynamical system, $\omega$ is its symplectic form  and $\H$ is its Hamiltonian. In particular, the notation $(T^*LK,\omega,\H_{\E_R})$ means that the cotangent bundle $T^*LK$ of the loop group $LK$ is the phase space of a non-linear $\sigma$-model living on the target manifold $K$, the symplectic form $\omega$ is the canonical one on the cotangent bundle and the Hamiltonian is $\H_{\E_R}$.  Of course, the triplet $(T^*LK,\omega,\H_{\E_R})$ is extracted from the action \eqref{mm1} of the $\sigma$-model  by the standard manipulations, which start  from defining the canonical momenta and end by writing up the first-order Hamiltonian  description of the dynamics.
 
Let us now identify all dynamical systems located at the corners of the  scheme \eqref{Sch2}. The lower-left corner $(T^*LK,\omega,\H_{\E_R})$ represent the   $\sigma$-model \eqref{mm1}, the upper-left corner
$(T^*LK,\omega,\H_{\E_L})$ is the $\sigma$-model \eqref{mm2}, the lower-right corner $(T^*L\tilde K_R,\tilde\omega_R,\tilde\H_{\E_R})$ corresponds to the action \eqref{mm3} and the upper-right corner
$(T^*L\tilde K_L,\tilde\omega_L,\tilde\H_{\E_L})$ to the action \eqref{mm4}. Actually, the target of the $\sigma$-model corresponding to the given corner can be always read off from the first entry of the triple, e.g.
$T^*LK$ means that the phase space of the $\sigma$-model is the cotangent bundle of the loop group $LK$, therefore the target of the corresponding $\sigma$-model is the group $K$.

The dynamical systems $(LD_L,\omega_{LD_L},Q_{\E_L})$
and $(LD_R,\omega_{LD_R},Q_{\E_R})$ located in the middle column of the scheme \eqref{Sch2} are the $\E$-models. In general, they live on non-isomorphic Drinfeld doubles $D_L\ne D_R$ but the scheme represents a non-trivial generalization of the Poisson-Lie T-duality even if $D_L$ is isomorphic to $D_R$ with only the subspaces $\E_L$ and $\E_R$ being different.

The scheme \eqref{Sch2} represents first of all the superposition of two standard Poisson-Lie dualities, one on the upper line of the scheme  and the other downstairs. Of course, if we want to apply the scheme to the context of the $\sigma$-models \eqref{mm1},\eqref{mm2},\eqref{mm3}  and \eqref{mm4}, 
we have to prove that the $\sigma$-model \eqref{mm1} is Poisson-Lie T-dual to the $\sigma$-model \eqref{mm3}
  via  some  intermediate  $\E$-model 
  $(LD_R,\omega_{LD_R},Q_{\E_R})$, while \eqref{mm2} is Poisson-Lie  T-dual to  \eqref{mm4} 
  via  some
  $(LD_L,\omega_{LD_L},Q_{\E_L})$. Indeed, after getting familiar with the properties of the affine (quasi-)Poisson groups in Sections 3 and 5, we shall be able to prove those facts in Sections 4 and 6. 
  
  Now we turn our attention to the up-down arrows in the scheme \eqref{Sch2} which indicate  that there exists the Hamiltonian-preserving symplectomorphism relating the $\E$-models  $(LD_L,\omega_{LD_L},Q_{\E_L})$ 
  and   $(LD_R,\omega_{LD_R},Q_{\E_R})$. 
  The existence of the non-trivial  symplectomorphisms $LD_L\to LD_R$ pulling back the Hamiltonian $Q_{\E_R}$   onto the Hamiltonian   $Q_{\E_L}$   is by no means an obvious thing, we shall nevertheless show in the present paper that such symplectomorphisms do exist for many choices of the Drinfeld doubles $\D_L,\D_R$. This fact leads to a substantial enlargement of the non-Abelian T-duality group as defined in \cite{LO} since the vertical arrow symplectomorphisms in the scheme \eqref{Sch2} makes possible to "travel" between the  Poisson-Lie T-dualities based on the different subspaces $\E_L\subset \D_L$ and $\E_R\subset \D_R$.  In particular, it follows from this that all four $\sigma$-models \eqref{mm1},\eqref{mm2},\eqref{mm3} and
  \eqref{mm4} are pairwise T-dual to each other. 
  
  Our procedure to prove the pairwise T-duality of the models \eqref{mm1},\eqref{mm2},\eqref{mm3} and
  \eqref{mm4} will be therefore as follows: given an affine (quasi-)Poisson group $K$, we find  the linear operators
$E^\na$, $E^{\ma}$, $\tilde E^\na$ and $\tilde E^{\ma}$ such that three things   hold: 1)
The $\sigma$-model \eqref{mm1} is Poisson-Lie T-dual to the $\sigma$-model \eqref{mm3}; 2)  the model \eqref{mm2}  is  Poisson-Lie T-dual to \eqref{mm4}; 3) there exists the symplectic automorphism of the symplectic manifold $(T^*LK,\omega)$ which pull backs the Hamiltonian $\H_{\E_R}$ to the Hamiltonian $\H_{\E_L}$. Note in this respect, that from the existence of all horizontal symplectomorphisms and of the left vertical symplectomorphism in the scheme \eqref{Sch2}, the existence of all remaining vertical
symplectomorphisms is automatically guaranteed. In particular, the right vertical symplectomorphisms exists and establishes the
affine (quasi-)Poisson T-duality relating the $\sigma$-models \eqref{mm3}  and \eqref{mm4}.

\section{Affine Poisson groups} 
 \setcounter{equation}{0}
 
The theory of the affine Poisson groups was developed in \cite{DS,Lu,K07} and the reader is either  invited to consult those references for finding  proofs of some
of the statements which we review in the present  section or they may extract them as the special cases of the full-fledged proofs given in Section 5  in the more general case of the affine quasi-Poisson groups. We mention also that some  parts of the theory  (like e.g. the  Poisson-Lie cohomology) are not included in this short review since they do not play an apparent  role in the T-duality story.

Let  $K$ be a Lie group equipped with a  Poisson bracket $\{f_1,f_2\}^\na$ defined for any pair of smooth functions $f_1,f_2$ on $K$ :
\be \{f_1,f_2\}^\na\equiv \la df_1\otimes df_2, \pi^\na\ra.\ee
The section $\pi^\na$ of the bundle $\Lambda^2TK$ is called the Poisson bivector corresponding to the    Poisson bracket $\{.,.\}^\na$.  The fact that the Poisson bracket $\{.,.\}^\na$ verifies the Jacobi identity is equivalent to the fact that  the  Schouten bracket of the bivector $\pi^\na$ with itself vanishes
\be [\pi^\na,\pi^\na]_K=0.\label{sch} \ee
Recall that the Schouten bracket $[.,.]_\K$ of bivectors is defined by the linear extension of the bracket  
\be [u_1\wedge u_2,v_1\wedge v_2]_K:=[u_1,v_1]_K\wedge u_2\wedge v_2+u_1\wedge [u_2,v_1]_K\wedge v_2- v_1\wedge [u_1,v_2]_K\wedge u_2-v_1\wedge u_1\wedge[u_2,v_2]_K,\label{247}\ee
where  $[u_1,v_1]_K$ stands for the standard Lie bracket of vector fields. 

The Poisson structure $\pi^\na$ on the group manifold $K$ is called {\it affine}, if the Lie derivatives of $\pi^\na$  with respect to the right-invariant and left-invariant vector fields satisfy, respectively
\be \L_{\nabla^l_{t_i}}\pi^\na=  -\jp\ ^{L}\tilde c_i^{\ jk}\nabla^l_{t_j}\wedge \nabla^l_{t_k};\label{pma}\ee
\be \L_{\nabla^r_{t_i}}\pi^\na=\jp \ ^{R}\tilde c_i^{\ jk}\nabla^r_{t_j}\wedge \nabla^r_{t_k}.\label{pmb}\ee
Here $t_i$ is some basis of the Lie algebra $\K$ of the group $K$, while $\ ^{L}\tilde c_i^{\ jk}$ and $\ ^{R}\tilde c_i^{\ jk}$
are, respectively, the structure constants of two Lie algebras denoted as $\tilde K_L$ and $\tilde K_R$. The quantities $\nabla^l_{t_j}$ and 
$\nabla^r_{t_j}$ are, respectively, the right-invariant and the left-invariant vector fields on $K$ which take value $t_i$ et the unit element $e_K$ of the group $K$. Explicitely, the vector fields $\nabla^l_{t_j}$ and
$\nabla^r_{t_j}$ act on the  functions on $K$ as
\be (\nabla^l_{t_i}f)(k)\equiv \frac{df(e^{st_i}k)}{ds}\bigg\vert_{s=0},\qquad (\nabla^r_{t_i}f)(k)\equiv \frac{df(ke^{st_i})}{ds}\bigg\vert_{s=0}.\ee
The crucial role in the theory of the affine Poisson groups is played by the value  of the bivector field $\pi^\na$ at the unit element $e_K$ of the group $K$.
We denote this value as $m$:
\be m:=\pi^\na(e_K)\equiv \jp m^{ij}t_i\wedge t_j.\ee 
Remark that $m$ is  naturally viewed as the element of $\Lambda^2\K$ where $\K$ is the Lie algebra of $K$. In the special case when  $m$ vanishes, the affine Poisson group  $(K,\pi^\na)$ is called the {\it Poisson-Lie group} and the finite-dimensional Lie algebras $\tilde\K_L$ and $\tilde\K_R^{opp}$ are isomorphic. If $m$ does not vanish, then $\pi^\na$ is not a Poisson-Lie structure on $K$
but it is true, however,  that {\it two different} Poisson-Lie structures $\pi^L$, $\pi^R$ on the group $K$ can be naturally constructed out of the affine Poisson structure $\pi^\na$. They are given by\footnote{Some overall signs in the definitions \eq{lr} and \eq{263} are conventional. In particular, we have flipped the signs of $\pi^{\ma}$ and of $\pi^R$ with respect to Ref. \cite{K07} in order to have the uniform positive sign in front of all the Poisson structures appearing in the formulae \eq{mm1},\eq{mm2},\eq{mm3} and \eq{mm4}.} the formulae
\be \pi^L=L_*m-\pi^\na\equiv \jp m^{ij}\nabla^r_{t_i}\wedge \nabla^r_{t_j}-\pi^\na, \qquad \pi^R=\pi^\na-R_*m \equiv \pi^\na-\jp m^{ij}\nabla^l_{t_i}\wedge \nabla^l_{t_j},\label{lr}\ee
where $L_*m$ and $R_*m$ are, respectively, the left- and right-invariant bivector fields on $K$ obtained by the left and right transport of the element $m$ to the whole group manifold. They verify
\be \L_{\nabla^l_{t_i}}\pi^L=  \jp\ ^{L}\tilde c_i^{\ jk}\nabla^l_{t_j}\wedge \nabla^l_{t_k};\label{266}\ee
\be \L_{\nabla^r_{t_i}}\pi^L=  \jp \ ^{L}\tilde c_i^{\ jk}\nabla^r_{t_j}\wedge \nabla^r_{t_k};\label{267}\ee
\be \L_{\nabla^l_{t_i}}\pi^R=  \jp\ ^{R}\tilde c_i^{\ jk}\nabla^l_{t_j}\wedge \nabla^l_{t_k};\label{268}\ee
\be \L_{\nabla^r_{t_i}}\pi^R= \jp \ ^{R}\tilde c_i^{\ jk}\nabla^r_{t_j}\wedge \nabla^r_{t_k}.\label{269}\ee

Note also, that the Poisson bivector $\pi^{\ma}$ given by the formula
\be \pi^{\ma}:= R_*m+L_*m-\pi^\na,\label{263}\ee
defines another affine Poisson structure on $K$ called the "mirror" one,
which satisfies
\be \L_{\nabla^l_{t_i}}\pi^{\ma}= -\jp\ ^{R}\tilde c_i^{\ jk}\nabla^l_{t_j}\wedge \nabla^l_{t_k};\label{270}\ee
\be \L_{\nabla^r_{t_i}}\pi^{\ma}=\jp \ ^{L}\tilde c_i^{\ jk}\nabla^r_{t_j}\wedge \nabla^r_{t_k}.\label{271}\ee
The two  Poisson-Lie structures associated to the mirror affine Poisson structure are the same as those associated to the original one  but their roles are reversed, the left one becomes the right one and vice versa. 

Drinfeld showed that each Poisson-Lie group $K$ of dimension $d$ can be embedded in a maximally isotropic way  as a subgroup in a $2d$-dimensional Lie group $D$ called the Drinfeld double of $K$. The "maximally isotropic" means that there is a symmetric non-degenerate ad-invariant bilinear form $(.,.)_\D$ defined on the Lie algebra $\D$ of $D$ which identically vanishes when restricted to  the Lie algebra $\K$ of the subgroup $K$.  
The structure of the  Drinfeld double "remembers" the Poisson-Lie bivector on $K$,  which  can be extracted   out of it from  the way how  certain another  maximally isotropic $d$-dimensional subgroup $\tilde K$ of $D$ is embedded in $D$. The role of the
groups $K$ and $\tilde K$ in the Poisson-Lie story is then interchangeable, which means that $\tilde K$ is also the  Poisson-Lie group and the Poisson-Lie bivector on $\tilde K$ can be again extracted from the way how $K$  and $\tilde K$ are embedded in the Drinfeld double.  In fact, any pair of maximally isotropic  embeddings of two $d$-dimensional subgroups $H$ and $\tilde H$   in $D$  defines the mutually dual Poisson-Lie structures on $H$ and $\tilde H$, provided that $\D$ can be written as the direct sum of  vector subspaces
Lie($H)\oplus$Lie($\tilde H$).

It follows that one can associate to every affine Poisson group $(K,\pi^\na)$ two Drinfeld doubles $D_L$ and $D_R$, which are, respectively, the Drinfeld doubles  of the Poisson-Lie groups $(K,\pi^L)$ and $(K,\pi^R)$.
The double $D_L$ has naturally two maximally isotropic subgroups $K$ and $\tilde K_L$ , while $D_R$ has subgroups $K$ and $\tilde K_R$. The Lie groups $\tilde K_L$ and $\tilde K_R$ are called  the left and the right dual groups of the affine Poisson group $K$ and their Lie algebras are isomorphic
to the Lie algebras $\tilde\K_L$ and $\tilde\K_R$ with the structure constants $\ ^{L}\tilde c_i^{\ jk}$ and $\ ^{R}\tilde c_i^{\ jk}$. 

For completeness, we detail how the left and right Poisson-Lie structures $\pi^L$ and $\pi^R$ as well as the mirror affine Poisson structure $\pi^{\ma}$ associated to the affine Poisson structure $\pi^\na$ are 
  extracted from the structure of the Drinfeld doubles of the affine Poisson group $K$ \cite{K07}. We start from the perspective of the double $D_L$,  
  we pick a basis $t_i, i=1,...,d$ of the Lie algebra $\K\subset\D_L$ and  the dual  basis $T^i_L, i=1,...,d$ of $\tilde\K_L\subset\D_L$   in such a way that 
\be (t_i,T^j_L)_{\D_L}=\delta_i^j.\label{20}\ee
We can then write
\be \ ^{L}\tilde c_i^{\ jk}=\left([T_L^j,T_L^k],t_i\right)_{\D_L}. \ee
The affine  Poisson bracket $\{.,.\}^{\na}_K$ corresponding   to the affine Poisson bivector $\pi^{\na}$ is then given by the formula\footnote{The formula \eq{pba} may suggest that adding any left-invariant $m$-term to a Poisson-Lie structure produces the affine Poisson structure but this is not true. In fact, the $m$-term has to be such that the affine Poisson structure be indeed Poisson, that is, the corresponding bracket of functions has to satisfy the Jacobi identity. An explicit condition for the good $m$-term can be written down but we do not need to know it for the T-duality story and the interested reader can find it e.g. in Ref.\cite{K07}.}
\be \{f_1,f_2\}_K^\na(k)=\jp m^{ij}\nabla^r_{t_i}f_1\nabla^r_{t_j}f_2+(T^i_L,Ad_kT^j_L)_{\D_L}\nabla^l_{t_i}f_1\nabla^r_{t_j}f_2;\label{pba}
\ee
and the Poisson-Lie bracket $\{.,.\}^{L}_K$ corresponding to the Poisson-Lie bivector $\pi^{L}$ is given by the formula
\be \{f_1,f_2\}_K^L(k)=-(T^i_L,Ad_kT^j_L)_{\D_L}\nabla^l_{t_i}f_1\nabla^r_{t_j}f_2.\label{plr}\ee
Here $f_1,f_2$ are smooth functions on $K$ and the Einstein summation convention applies.

From the point of view of the double $D_R$,  we pick a basis $t_i, i=1,...,d$ of the Lie algebra $\K\subset\D_R$ and the dual basis $T^i_R, i=1,...,d$ of $\tilde\K_R\subset\D_R$ in such a way that 
\be (t_i,T^j_R)_{\D_R}=\delta_i^j.\label{308}\ee
We can then write
\be \ ^{R}\tilde c_i^{\ jk}=\left([T_R^j,T_R^k],t_i\right)_{\D_R}.\ee
The mirror affine  Poisson bracket $\{.,.\}^{\ma}_K$ corresponding  to the mirror affine Poisson bivector $\pi^{\ma}$ is  then given by the formula
\be \{f_1,f_2\}_K^{\ma}(k)=\jp m^{ij}\nabla^r_{t_i}f_1\nabla^r_{t_j}f_2+(T^i_R,Ad_kT^j_R)_{\D_R}\nabla^l_{t_i}f_1\nabla^r_{t_j}f_2\label{pbam}
\ee
and the Poisson bracket $\{.,.\}^{R}_K$ corresponding to the Poisson-Lie bivector $\pi^{R}$ is  given by  
\be   \{f_1,f_2\}_K^R(k)=-(T^i_R,Ad_kT^j_R)_{\D_R}\nabla^l_{t_i}f_1\nabla^r_{t_j}f_2.\label{314}\ee
 
  We finish this section by proving the following useful relation
 \be \{f_1\circ S,f_2\circ S\}_K^\na=\{f_1,f_2\}_K^{\ma}\circ S,\label{331}\ee
 where $S:K\to K$ is defined as $S(k):=k^{-1}$.
 First of all, we show that   for the Poisson-Lie structure $\pi^L$ given by the formula \eq{plr} it holds
  \be \{f_1\circ S,f_2\circ S\}_K^L=-\{f_1,f_2\}_K^L\circ S.\label{334}\ee
 This follows from the repeated use of the following obvious identity
  \be \nabla^l_{t_i}(f\circ S)=-\left(\nabla^r_{t_i}f\right)\circ S,\label{336}\ee
indeed, we have
$$ \{f_1\circ S,f_2\circ S\}^L_K(k)=-(T^i_L,Ad_kT^j_L)_{\D_L}\nabla^l_{t_i}(f_1\circ S)\nabla^r_{t_j}(f_2\circ S)=$$\be = -(Ad_{k^{-1}}T^i_L,T^j_L)_{\D_L}\left((\nabla^r_{t_i}f_1)\circ S\right)\left((\nabla^l_{t_j}f_2)\circ S\right)= -\{f_1,f_2\}_K^L(k^{-1}).\ee
Recall that we have $\pi^{\ma}=\pi^L+R_*m$ and $\pi^\na=-\pi^L+L_*m$, therefore for proving \eq{331}, we have to show
that 
\be m^{jk}
\nabla^l_{t_j}(f_1\circ S)\wedge \nabla^l_{t_k}(f_2\circ S)=m^{jk}
\left(\nabla^r_{t_j}f_1\right)\circ S \left(\nabla^r_{t_k}f_1\right)\circ S.\ee
But this follows from Eq.\eq{336}.

\section{Affine Poisson T-duality}
 \setcounter{equation}{0}
 
The strategy of our presentation in the present section  is  as follows: we  first  assume that  the T-duality between the models \eqref{mm1}  and \eqref{mm2} indeed holds and then  we infer from this assumption that all $\sigma$-models \eqref{mm1},\eqref{mm2},\eqref{mm3}  and \eqref{mm4} are pairwise T-dual
to each other and fit into the scheme \eqref{Sch2}. In  subsection 4.2, we formulate a sufficient condition to be fulfilled by the data $K,\Pi^{\na},E^{\na},E^{\ma}$  in order that the assumed duality between  \eqref{mm1}  and \eqref{mm2} really takes place, and, in Subsection 4.3, we  construct  a large class of examples of the affine Poisson groups and of the operators  $E^{\na}$ and $E^{\ma}$ for which this sufficient condition is satisfied.

\subsection{Pairwise T-duality of four $\sigma$-models}

  Let  $\tilde K_L$ and $\tilde K_R$ be the dual groups of a given affine Poisson group 
 $(K,\pi^\na)$
and let $D_L$ and $D_R$ be their respective Drinfeld doubles. Consider the $\sigma$-models \eqref{mm1} and  \eqref{mm2} living on $K$:
\be S_{\na}(k)=\jp\int d\tau \oint \biggl(\Bigl(E^{\na}+\Pi^{\na}(k)\Bigr)^{-1}\partial_+k k^{-1}, \partial_- k k^{-1}\biggr)_{\D_R},\label{af1}\ee
\be S_{\ma}(k)=\jp\int d\tau \oint \biggl(\Bigl(E^{\ma}+\Pi^{ \ma}(k)\Bigr)^{-1}\partial_+k k^{-1}, \partial_- k k^{-1}\biggr)_{\D_L}.\label{af2}\ee
where $E^{\na}:\tilde\K_R\to\K$ and $E^{\ma}:\tilde\K_L\to\K$ are some invertible linear operators, and  $\Pi^{\na}(k):\tilde\K_R\to\K$ and $\Pi^{ \ma}(k):\tilde\K_L\to\K$
are, respectively, the affine Poisson structure and the mirror affine Poisson structure on $K$ in the sense of the formulae
\be \{f_1,f_2\}^{\na}_K=\left(\ \!^R\nabla^l f_1,\Pi^{\na}(k)\ \!^R\nabla^l f_2\right)_{\D_R},\label{mor1}\ee
\be \{f_1,f_2\}^{\ma}_K=\left(\ \!^L\nabla^l f_1,\Pi^{\ma}( k)\ \!^L\nabla^l f_2\right)_{\D_L}.\label{mor2}\ee
Here $\ \!^R\nabla^l$ and $\ \!^L\nabla^l$ are, respectively, $\tilde\K_R$- and $\tilde\K_L$-valued differential operators acting on the functions on $K$  as
\be \left(\ \!^{L}\nabla^l f, x\right)_{\D_L}:= \nabla^l_{x} f,\quad \left(\ \!^{R}\nabla^l f, x\right)_{\D_R}:= \nabla^l_{x} f, \qquad  x\in \K.\label{sab}\ee

 \noindent {\bf Assertion 1}: {\it If the $\sigma$-model \eqref{af1} is dual to the $\sigma$-model \eqref{af2} for some choice of the linear operators} $E^\na$ {\it and} $E^{\ma}$, {\it  then there exist operators} $\tilde E^{\na}$ {\it  and} $\tilde E^{\ma}$ {\it such that  all $\sigma$-models \eqref{mm1},\eqref{mm2},\eqref{mm3}  and \eqref{mm4} are pairwise T-dual
to each other.}
 
 \medskip
 
Let us prove the {\bf Assertion 1}. We set 
 \be M^R=\Pi^\na(e_K), \quad  M^L=\Pi^{\ma}(e_K)\label{mlr'}\ee
 and rewrite the pair of the actions \eqref{af1}, \eqref{af2}   equivalently as
 \be S_\na(k)=\jp\int d\tau \oint \biggl(\Bigl(E^\na+M^R+\Pi^R(k)\Bigr)^{-1}\partial_+k k^{-1}, \partial_- k k^{-1}\biggr)_{\D_R},\label{eadl}\ee
\be S_{\ma}(k)=\jp\int d\tau \oint \biggl(\Bigl(E^{\ma}+M^L+\Pi^L(k)\Bigr)^{-1}\partial_+k k^{-1}, \partial_- k k^{-1}\biggr)_{\D_L},\label{eadr}\ee
where $\Pi^L(k):\tilde\K_L\to\K$ and $\Pi^R(k):\tilde\K_R\to\K$ stand, respectively, for the left and right Poisson-Lie structures on $K$, associated to the affine Poisson structure in the sense of the formulae
\be \{f_1,f_2\}^L_K=\left(\ \!^L\nabla^l f_1,\Pi^L(k)\ \!^L\nabla^lf_2\right)_{\D_L},\label{mor11}\ee
\be \{f_1,f_2\}^{R}_K=\left(\ \!^R\nabla^l f_1,\Pi^R( k)\ \!^R\nabla^l f_2\right)_{\D_R}.\label{mor22}\ee
The crux of the proof of the {\bf Assertion 1} is now obvious, because the model \eqref{af1} rewritten
as \eqref{eadl} is of the type \eqref{ea}, hence it is Poisson-Lie T-dualizable. Similarly, the model \eqref{af2} rewritten
as \eqref{eadr} is also of the type \eqref{ea}, hence it is also Poisson-Lie T-dualizable. The standard Poisson-Lie T-duality  reviewed in Section 2 then gives, respectively, the dual models  of  \eqref{eadl} and of \eqref{eadr}:
\be \tilde S_{\na}(\tilde k_R)=\jp\int d\tau \oint \biggl(\Bigl(\tilde E^{\na}+\tilde\Pi^R(\tilde k_R)\Bigr)^{-1}\partial_+\tilde k_R\tilde k_R^{-1}, \partial_-\tilde k_R\tilde k_R^{-1}\biggr)_{\D_R},\label{ear''}\ee
\be \tilde S_{\ma}(\tilde k_L)=\jp\int d\tau \oint \biggl(\Bigl(\tilde E^{\ma}+\tilde\Pi^L(\tilde k_L)\Bigr)^{-1}\partial_+\tilde k_L\tilde k_L^{-1}, \partial_-\tilde k_L\tilde k_L^{-1}\biggr)_{\D_L}.\label{eal''}\ee
 Here $\tilde k_L,\tilde k_R$ are, respectively, $\tilde K_L$,$\tilde K_R$-valued $\sigma$-model fields, $\tilde\Pi^L(\tilde k_L)$, $\tilde\Pi^R(\tilde k_R)$ are the Poisson-Lie structures on $\tilde K_L,\tilde K_R$ dual to the Poisson-Lie structures $\Pi^L(k)$, $\Pi^R(k)$ on $K$,  and   $\tilde E^\na$, $\tilde E^{\ma}$ are the inverse  operators to 
 $E^\na+M^R$ and to $E^{\ma}+M^L$. The models \eqref{ear''} and \eqref{eal''} coincide with the models \eqref{mm3} and \eqref{mm4},
 therefore we conclude that all $\sigma$-models \eqref{mm1},\eqref{mm2},\eqref{mm3}  and \eqref{mm4} are pairwise T-dual
to each other. We note, moreover, that the subspaces $\E_R$ and $\E_L$ featuring in the scheme \eqref{Sch2} are, respectively 
   $\E_L=\{\tilde x +(E^{\ma}+M^L)\tilde x,\tilde x\in\tilde\K_L\}$ and  $\E_R=\{\tilde x +(E^{\na}+M^R)\tilde x,\tilde x\in\tilde\K_R\}$.  
 
 \subsection{Sufficient condition for the affine Poisson T-duality}

 So far we have established, that all $\sigma$-models \eqref{mm1},\eqref{mm2},\eqref{mm3}  and \eqref{mm4} are pairwise T-dual
to each other if we succeed to associate a symplectomorphism to the  left  vertical arrow in the scheme \eqref{Sch2}, or, in other words, if we prove that the models \eqref{mm1} and \eqref{mm2} are T-dual to each other. 
The principal result of the present subsection is the {\bf Assertion 2} which states  the sufficient condition on the 
 data $K,\Pi^\na,E^\na,E^{\ma}$ guaranteeing  the existence of the seeken left vertical symplectomorphism.
 In order to formulate this condition, we need to define two bilinear forms $(.,.)_\na$ and $(.,.)_{\ma}$  on the Lie algebra $\K$:
 \be (x,y)_\na:=(x,(E^\na)^{-1}y)_{\D_R},\quad (x,y)_{\ma}:=(x,(E^{\ma})^{-1}y)_{\D_L},\quad x,y\in \K.\label{lem}\ee
 
 \medskip 
 
 \noindent {\bf Assertion 2}: {\it If the bilinear form} $(.,.)_\na$ {\it is symmetric, non-degenerate and 
 ad$_\K$-invariant and, moreover, if it coincides with the bilinear form} $(.,.)_{\ma}$ {\it then the 
 $\sigma$-model \eqref{mm1} is T-dual to the $\sigma$-model \eqref{mm2}. } 
 
 \medskip
 
 In order to prove {\bf Assertion 2}, we first remark that it holds
 \be E^\na\ \!^R\nabla^l=E^{\ma}\ \!^L\nabla^l=\ \!^\na{\bm\nabla}^l,\label{plk}\ee
 where the operators $\ \!^R\nabla^l$ and $\ \!^L\nabla^l$ where defined in \eqref{sab} and we define 
a $\K$-valued differential operator $\ \!^\na{\bm\nabla}^l$ acting on the functions on $K$ as
 \be \left(\ \!^\na{\bm\nabla}^l f, x\right)_\na := \nabla^l_{x} f, \qquad  x\in \K.\label{bfs}\ee
 To see e.g. that the first of the relations \eqref{plk} indeed holds, we rewrite the left-hand-side of the second of Eqs. \eqref{sab}
 as 
 \be \left(\ \!^\na{\bm\nabla}^l f, x\right)_\na := \nabla^l_{x} f = \left(\ \!^{R}\nabla^l f, x\right)_{\D_R}=\left({E^\na}^{-1}E^\na\ \!^{R}\nabla^l f, x\right)_{\D_R}=
 \left(E^\na\ \!^{R}\nabla^l f, x\right)_\na \ee
 and  we finish up the argument by invoking the non-degeneracy of the bilinear forms $(.,.)_{\D_R}$ and $(.,.)_\K$.
 
 Our next goal is to prove the following relation
 \be \Pi^{\ma}(k)= Ad_k\Pi^\na(k^{-1})({E^\na})^{-1}Ad_{k^{-1}}E^{\ma},\label{cru}\ee
 where the operators $\Pi^{\ma}(k)$ and $\Pi^{\na}(k)$ where defined respectively in Eqs. \eqref{mor1} and \eqref{mor2}.
 To prove \eqref{cru}, we employ the identity \eq{331}. By using the relations \eqref{plk}, we can rewrite the affine Poisson brackets \eqref{mor1} and \eqref{mor2}
 as 
 \be \{f_1,f_2\}^\na_K=\left({E^\na}^{-1}E^\na\ \!^R\nabla^l f_1,\Pi^\na(k){E^\na}^{-1}E^\na\ \!^R\nabla^l f_2\right)_{\D_R}=
 \left(  \ \!^\na{\bm\nabla}^l f_1,\Pi^\na(k){E^\na}^{-1}  \ \!^\na{\bm\nabla}^l f_2\right)_\na, \label{blb1}\ee
 and, similarly,
  \be \{f_1,f_2\}^{\ma}_K= 
 \left(  \ \!^\na{\bm\nabla}^l f_1,\Pi^{\ma}(k){E^{\ma}}^{-1}  \ \!^\na{\bm\nabla}^l f_2\right)_{\ma}.\label{blb2} \ee
 We  obtain the desired identity \eqref{cru} from the relations \eqref{331}, \eqref{blb1}, \eqref{blb2} and also from the following identity
 \be Ad_k(  \ \!^\na{\bm\nabla}^rf(k))=  \ \!^\na{\bm\nabla}^lf(k).\ee
 Let us  rewrite the $\sigma$-model actions \eqref{mm1} and \eqref{mm2} as follows
$$ S_\na(k)= \jp\int d\tau \oint \biggl({E^\na}^{-1}\Bigl(1+\Pi^\na(k){E^\na}^{-1}\Bigr)^{-1}\partial_+k k^{-1}, \partial_- k k^{-1}\biggr)_{\D_R} =$$\be =\jp\int d\tau \oint \biggl(\Bigl(1+\Pi^\na(k){E^\na}^{-1}\Bigr)^{-1}\partial_+k k^{-1}, \partial_- k k^{-1}\biggr)_\na\label{e1}\ee
 $$ S_{\ma}(k)= \jp\int d\tau \oint \biggl({E^{\ma}}^{-1}\Bigl(1+\Pi^{\ma}(k){E^{\ma}}^{-1}\Bigr)^{-1}\partial_+k k^{-1}, \partial_- k k^{-1}\biggr)_{\D_L} =$$\be =\jp\int d\tau \oint \biggl(\Bigl(1+\Pi^{\ma}(k){E^{\ma}}^{-1}\Bigr)^{-1}\partial_+k k^{-1}, \partial_- k k^{-1}\biggr)_{\ma}.\label{e2}\ee
 Using the crucial identity \eqref{cru} as well as the hypothesis that the ad$_\K$-invariant bilinear forms   $(.,.)_\na$ and $(.,.)_{\ma}$ coincide, we find finally
 \be S_\na(k)=S_{\ma}(k^{-1}).\label{ama}\ee
 The equation \eqref{ama} may look surprising because it proves the T-duality between the $\sigma$-models 
 \eqref{mm1} and \eqref{mm2} in the simplest possible way; indeed, the model  \eqref{mm1} becomes  the model \eqref{mm2}
 by the simple field redefinition $k\to k^{-1}$ and the corresponding T-duality symplectomorphism is therefore just the so called "point canonical transformation". In the standard T-duality story, the point canonical transformation is normally  not considered to be an interesting one since it establishes the dynamical equivalence of two $\sigma$-models just by field redefinitions, so what is the point in discovering that 
 the left vertical symplectomorphism in the scheme \eqref{Sch2} relating the models \eqref{mm1} and  \eqref{mm2} is just the point canonical transformation? In fact, the crux of the affine Poisson T-duality is the   {\it nontriviality of the right  vertical symplectomorphism} relating the $\sigma$-models  \eqref{mm3} and \eqref{mm4}.  This right  vertical symplectomorphism is not the point canonical transformation since it is the composition
 of the upper horizontal, left  vertical and lower horizontal symplectomorphisms appearing in  the scheme \eqref{Sch2} and the horizontal symplectomorphisms, being the Poisson-Lie T-dualities, are not the point canonical transformations. 
 
 Summarizing: if the sufficient condition formulated in {\bf Assertion 2} holds  then there is the nontrivial
 T-duality transformation relating the $\sigma$-models living on the dual groups $\tilde K_L$ and $\tilde K_R$
 of the given affine Poisson group $K$:
  \be \tilde S_{\na}(\tilde k_R)=\jp\int d\tau \oint \biggl(\Bigl(\tilde E^{\na}+\tilde\Pi^R(\tilde k_R)\Bigr)^{-1}\partial_+\tilde k_R\tilde k_R^{-1}, \partial_-\tilde k_R\tilde k_R^{-1}\biggr)_{\D_R}.\label{eal'}\ee
 \be \tilde S_{\ma}(\tilde k_L)=\jp\int d\tau \oint \biggl(\Bigl(\tilde E^{\ma}+\tilde\Pi^L(\tilde k_L)\Bigr)^{-1}\partial_+\tilde k_L\tilde k_L^{-1}, \partial_-\tilde k_L\tilde k_L^{-1}\biggr)_{\D_L}.\label{ear'}\ee

 This is the principal statement of the affine Poisson T-duality story.
 
 \subsection{Examples of the affine Poisson T-duality}

 If an  affine Poisson group $(K,\pi^\na)$ together with some operators $E^\na,E^{\ma}$ satisfy the 
  sufficient condition of {\bf Assertion 2} then the Lie algebra $\K$ is quadratic, which means that  there is a symmetric ad$_\K$-invariant non-degenerate
  bilinear form $(.,.)_\K$ on  $\K$.  Moreover, we restrict  our attention to such affine Poisson structures $\pi^\na$ on the quadratic\footnote{A Lie group $K$ is said quadratic if its Lie algebra $\K$ is quadratic, that is, if it exists an ad-invariant symmetric non-degenerate bilinear form on $\K$.}  group $K$
  for which the associated  Drinfeld doubles $D_L$ and $D_R$ are isomorphic as the Lie groups, and both Lie algebras $\D_L$ and $\D_R$ have the structure of the Lie  algebra $\D_\e$ defined as follows: the elements of $\D_\e$  are pairs
  $(x_1,x_2)$, $x_1,x_2\in\K$ and  the Lie algebra commutator is given by the expression
\be [(x_1,x_2),(y_1,y_2)]_\e=([x_1,y_1]+\e[x_2,y_2],[x_1,y_2]+[x_2,y_1]).\label{str}\ee
The symmetric non-degenerate  ad-invariant  bilinear form $(.,.)_{\D_\e}$ on $\D_\e$ does not depend on the real negative parameter $\e$ and it is given by the
formula
\be ((x_1,x_2),(y_1,y_2))_{\D_\e}:= (x_2,y_1)_\K+(x_1,y_2)_\K.\label{bil}\ee
The  Lie algebra $\K$ is embedded into $\D_\e$ in the maximally isotropic way as $(\K,0)\subset \D_\e$.

As we already know, the affine Poisson structure $\pi^\na$ induces two maximally isotropic Lie subalgebras $\tilde\K_L$, $\tilde\K_R$ of the respective doubles $\D_L$ and $\D_R$.  Since we consider the case $\D_L=\D_R=\D_\e$,
the both Lie algebras  $\tilde\K_L$ and $\tilde\K_R$  must be maximally isotropic subalgebras of $\D_\e$.
Recall that $\D_\e$ as the vector space (but not necessarily as the Lie algebra) can be  written in terms of the direct sums \be\D_\e=\K\oplus \tilde\K_L=\K\oplus\tilde\K_R.\label{dso}\ee
This fact as well as  the structure of the  Lie bracket \eqref{str} on the double $\D_\e$  entail the existence of two linear operators $R^r, R^l:\K\to\K$ such that the Lie subalgebras $\tilde\K_L,\tilde K_R\subset \D_\e$ can be written as the following  graphs
\be \tilde\K_L =\{(-R^lx,x),x\in\K\},\quad  \tilde\K_R =\{(-R^rx,x),x\in\K\}.\label{max}\ee
 Moreover, the fact that the restrictions of the bilinear form $(.,.)_\D$ on the both $\tilde\K_L$ and $\tilde K_R$ must   vanish implies the skew-symmetry of the both operators $R^r,R^l$:
 \be (R^rx,y)_\K=-(x,R^ry)_\K,\quad  (R^lx,y)_\K=-(x,R^ly)_\K,   \qquad x,y\in\K.\ee
 Finally, the fact that $\tilde\K_L$ and $\tilde K_R$ are the  Lie subalgebras of $\D_\e$ sets the following constraints on the operators $R^r$ and $R^l$:
 \be [R^rx,R^ry]=R^r([R^rx,y]+[x,R^ry])-\e[x,y], \qquad \forall x,y\in \K, \label{YBr}\ee 
  \be [R^lx,R^ly]=R^l([R^lx,y]+[x,R^ly])-\e[x,y], \qquad \forall x,y\in \K, \label{YBl}\ee 
 The constraints \eqref{YBr} and \eqref{YBl} are known under the name of the Yang-Baxter equations.
 
 Actually, the knowledge of two skew-symmetric solutions  $R^r$ and $R^l$ of the Yang-Baxter equation is
 all what is needed for reconstructing the affine Poisson structure $\pi^\na$ on the quadratic Lie  group $K$.
 The affine Poisson bracket $\{.,.\}^\na$, its mirror  $\{.,.\}^{\ma}$ and the associated left and write Poisson-Lie brackets  $\{.,.\}^L$,  $\{.,.\}^R$ can be easily extracted from Eqs. \eqref{pba}, \eqref{pbam} and \eqref{max} and they are given by the formulae
 \be \{f_1,f_2\}^\na_K:=({\bm\nabla}^r f_1,R^r{\bm\nabla}^r f_2)_\K+({\bm\nabla}^l f_1,R^l{\bm\nabla}^l f_2)_\K.\label{aff}\ee
\be \{f_1,f_2\}^{\ma}_K:=({\bm\nabla}^r f_1,R^l{\bm\nabla}^r f_2)_\K+({\bm\nabla}^l f_1,R^r{\bm\nabla}^l f_2)_\K,\label{afg}\ee
\be \{f_1,f_2\}^L_K:=({\bm\nabla}^r f_1,R^l{\bm\nabla}^r f_2)_\K-({\bm\nabla}^l f_1,R^l{\bm\nabla}^l f_2)_\K,\label{l}\ee
\be \{f_1,f_2\}^{R}_K:=({\bm\nabla}^r f_1,R^r{\bm\nabla}^r f_2)_\K-({\bm\nabla}^l f_1,R^r{\bm\nabla}^l f_2)_\K.\label{r}\ee
Here we have defined 
 the $\K$-valued differential operators ${\bm\nabla}^r, {\bm\nabla}^l$ acting on the functions on $K$ as
 \be  \left({\bm\nabla}^r f, x\right)_\K:=  \nabla^r_{x} f   \qquad
  \left({\bm\nabla}^l f, x\right)_\K:= \nabla^l_{x} f  , \qquad  x\in \K.\label{bft'}\ee
Working with the affine Poisson structure given by Eq. \eqref{aff}, can we find the operators 
$E^\na:\tilde\K_R\to\K$ and $E^{\ma}:\tilde\K_L\to\K$ which would fulfil the sufficient condition of {\bf Assertion 2}? The answer to this question is affirmative;  in the two cases, it is given by:
\be E^\na(-R^rx,x):=a(x,0), \qquad  E^{\ma}(-R^lx,x):=a(x,0), \qquad a<0.\label{ho}\ee
Using the definition \eqref{bil},  it is then easy to verify for every $x,y\in \K$ that it holds
\be (x,y)_\na\equiv ((x,0),(E^\na)^{-1}(y,0))_\D=\frac{1}{a}((x,0),(-R^ry,y))_\D=\frac{1}{a}(x,y)_\K,\ee
\be (x,y)_{\ma}\equiv ((x,0),(E^{\ma})^{-1}(y,0))_\D=\frac{1}{a}((x,0),(-R^ly,y))_\D=\frac{1}{a}(x,y)_\K,  \ee
hence the sufficient condition for the affine Poisson T-duality is indeed satisfied.

It is  instructive to cast the $\sigma$-model actions \eqref{mm1} and \eqref{mm2} in terms of the Yang-Baxter operators $R^l,R^r$. For that, we combine the
formulae \eqref{blb1}, \eqref{blb2} with \eqref{aff}, \eqref{afg} to find
\be \Pi^\na(k)({E^\na})^{-1}=\frac{1}{a}R^l+\frac{1}{a}R^r_{k^{-1}}, \quad \Pi^{\ma}(k)({E^{\ma}})^{-1}=\frac{1}{a}R^r+\frac{1}{a}R^l_{k^{-1}},\label{ger}\ee
where 
\be R^l_{k^{-1}}:=Ad_k R^l Ad_{k^{-1}}, \quad   R^r_{k^{-1}}:=Ad_k R^r Ad_{k^{-1}}.\ee 
Using then the formulae \eqref{e1}, \eqref{e2} and \eqref{ger}, we infer
\be S_\na(k)=\jp\int d\tau \oint \left(\left(a+ R^l+R^r_{k^{-1}}\right)^{-1}\partial_+k k^{-1}, \partial_- k k^{-1}\right)_\K,\label{kad}\ee
\be S_{\ma}(k)=\jp\int d\tau \oint \left(\left(a+R^r+R^l_{k^{-1}}\right)^{-1}\partial_+k k^{-1}, \partial_- k k^{-1}\right)_\K,\label{kbd}\ee
There is no universal way to rewrite the actions of the dual $\sigma$-models $\tilde S_\na(\tilde k_R)$ and $\tilde S_{\ma}(\tilde k_L)$
in terms of the operators $R^r,R^l$ since the very structure of the groups $\tilde K_L,\tilde K_R$ depends implicitely  on $R^r,R^l$. On the other hand, the operators $R^r$, $R^l$ are very useful if we wish to describe explicitely the subspaces $\E_L$, $\E_R$ underlying, respectively, the Poisson-Lie T-dualities relating the models \eqref{mm1} with \eqref{mm3} and \eqref{mm2} with \eqref{mm4}. We find with the help of the formulae  \eqref{mlr'}, \eqref{ho} and \eqref{ger} that
\be M^R(E^\na)^{-1}= M^L(E^{\ma})^{-1}=\frac{1}{a}R^l+\frac{1}{a}R^r,\label{mlr}\ee
hence
 \be \E_R=\{\tilde x_R +(E^\na+M^R)\tilde x_R,\tilde x_R\in\tilde\K_R\}=\{(E^\na)^{-1}x +(E^\na+M^R)(E^\na)^{-1}x , x\in\K\}=\{(ay+R^ly,y),y\in\K\},\label{ell}\ee
 \be \E_L=\{\tilde x_L +(E^{\ma}+M^L)\tilde x_L,\tilde x_L\in\tilde\K_L\}=\{(E^{\ma})^{-1}x +(E^{\ma}+M^L)(E^{\ma})^{-1}x , x\in\K\}=\{(ay+R^ry,y),x\in\K\}.\label{err}\ee
Note that as far as $R^r\neq R^l$, the subspaces $\E_L$ and $\E_R$ do not coincide and the scheme \eqref{Sch2} hence represents the generalization of the Poisson-Lie T-duality (recall that the Poisson-Lie T-duality is based on the unique subspace). In the special case when $R^l=R^r$,  the dual
groups $\tilde K^L$ and $\tilde K^R$ coincide  as well as do the subspaces $\E_L$ and $\E_R$. This means that the $\sigma$-models \eqref{mm3} and \eqref{mm4} coincide too
and the upper line of the affine Poisson T-duality scheme \eqref{Sch2} merges in this case with the lower line to become the Poisson-Lie T-duality scheme \eqref{Sch1'}. In conclusion, if $R^r=R^l$,
the affine Poisson T-duality becomes the standard Poisson-Lie T-duality.

\subsection{The affine Poisson T-duality and the Drinfeld twist operators}

In this subsection, we choose for $\K$ a compact simple Lie algebra equipped with its Killing-Cartan form $(.,.)_\K$ and for its Drinfeld double  we take $\D_\e$ for $\e=0$ (cf. Eq.\eqref{str}). We now pick a Cartan subalgebra $\gH\in\K$ and we consider the subspace $\gH^\perp\subset \K$ which is perpendicular to $\gH$ with respect to the Killing-Cartan form $(.,.)_\K$. 
We shall call any skew-symmetric operator $R:\K\to\K$   the Drinfeld twist operator, if $\gH^\perp\subset {\rm Ker}(R)$ and Im$(R)\subset\gH$. Any Drinfeld twist operator verifies automaticallly the Yang-Baxter condition  \eqref{YBl} because of the commutativity of the Cartan subalgebra. In the case when the affine Poisson structure \eqref{aff} is given by the Drinfeld twist operators $R^l,R^r$, the $\sigma$-models \eqref{eal'} and \eqref{ear'} dual to the models \eqref{kad} and \eqref{kbd} can be written more explicitly and  we  present here  the corresponding formulae.

We start with the description of the Lie group  $D_0$ integrating the Lie algebra $\D_0$ (the commutator of $\D_0$ is given by Eq.\eqref{str} for $\e=0$). The element of $D_0$ are the pairs $(k,\kappa)$, where $k\in K$ and $\kappa\in\K$, the group multiplication in $D_0$ is given by \be (k_1,\kappa_1)(k_2,\kappa_2)=(k_1k_2,\kappa_1+Ad_{k_1}\kappa_2)\label{gl}\ee
and the inverse element by
\be (k,\kappa)^{-1}=(k^{-1},-Ad_{k^{-1}}\kappa).\label{ind}\ee
In what follows, we shall moreover need  explicit expressions for the Maurer-Cartan forms on $D_0$ as well  as those for the adjoint action of the group $D_0$ on the Lie algebra $\D_0$. We have, respectively,  for the left- and right-invariant forms
\be (k,\kappa)^{-1}d(k,\kappa)=\left(k^{-1}dk,Ad_{k^{-1}}(  d\kappa)\right),\quad d(k,\kappa)(k,\kappa)^{-1}=\left(dkk^{-1},d\kappa+[\kappa,dkk^{-1}]\right)\label{dif} \ee
and, for the adjoint action
\be Ad_{(k,\kappa)}(x_1,x_2)= (Ad_kx_1,Ad_kx_2+[\kappa,Ad_kx_1]).\label{adj}\ee
The elements of the subgroup $K$ of $D_0$ have the form $(k,0)$ and the elements of the subgroups $\tilde K_L$  and $\tilde K_R$ have, respectively, the form
\be \tilde K_L=\{ (e^{-R^l\kappa},\kappa),\kappa\in\K\},\quad \tilde  K_R=\{ (e^{-R^r\kappa},\kappa),\kappa\in\K\}.\label{dgs}\ee
The reader may verify by direct computations, that the definitions \eqref{dgs} yield  the Poisson brackets
\eqref{aff}, \eqref{afg}, \eqref{l} and \eqref{r} via Eqs. \eqref{pba}, \eqref{plr}, \eq{pbam} and \eq{314}.

In order to write down explicitly the actions \eqref{eal'} and\eqref{ear'}, we first represent them in the following form
\be \tilde S_{\na}=\jp\int d\tau \oint d\sigma \biggl(\Bigl(1+(E^\na+M^R)\tilde\Pi^R(\tilde k_R)\Bigr)^{-1}(E^\na+M^R)\partial_+\tilde k_R\tilde k_R^{-1}, \partial_-\tilde k_R\tilde k_R^{-1}\biggr)_{\D_R}.\label{ealb}\ee
 \be \tilde S_{\ma}=\jp\int d\tau \oint d\sigma \biggl(\Bigl(1+(E^{\ma}+M^L)\tilde\Pi^L(\tilde k_L)\Bigr)^{-1}(E^{\ma}+M^L)\partial_+\tilde k_L\tilde k_L^{-1}, \partial_-\tilde k_L\tilde k_L^{-1}\biggr)_{\D_L}.\label{earb}\ee 
We find from the formulae \eqref{dif} and \eqref{dgs} 
\be \partial_\pm\tilde k_L \tilde k_L^{-1}=\left(-R^l \partial_\pm\kappa, \partial_\pm\kappa -[\kappa,R^l \partial_\pm\kappa]\right),\qquad \partial_\pm\tilde k_R \tilde k_R^{-1}=\left(-R^r \partial_\pm\kappa, \partial_\pm\kappa -[\kappa,R^r \partial_\pm\kappa]\right)\label{605}\ee
and from the formulae \eqref{ho} and  \eqref{mlr}, we infer
\be (E^\na+M^R)\partial_\pm\tilde k_R\tilde k_R^{-1}=\left((a+R^l+R^r)(\partial_\pm\kappa -[\kappa,R^r \partial_\pm\kappa]),0\right);\ee
\be (E^{\ma}+M^L)\partial_\pm\tilde k_L\tilde k_L^{-1}=\left((a+R^l+R^r)(\partial_\pm\kappa -[\kappa,R^l \partial_\pm\kappa]),0\right).\ee
Thus, taking into account the formula \eqref{bil}, we can rewrite Eqs. \eqref{ealb} and \eqref{earb} in terms of the bilinear form $(.,.)_\K$.
\be \tilde S_{\na}=\jp\int d\tau \oint d\sigma \biggl(\Bigl(1+(E^\na+M^R)\tilde\Pi^R(\tilde k_R)\Bigr)^{-1}(a+R^l+R^r)(\partial_+\kappa -[\kappa,R^r \partial_+\kappa]), \partial_-\kappa -[\kappa,R^r \partial_-\kappa]\biggr)_\K.\label{ealb'}\ee
 \be \tilde S_{\ma}=\jp\int d\tau \oint d\sigma \biggl(\Bigl(1+(E^{\ma}+M^L)\tilde\Pi^L(\tilde k_L)\Bigr)^{-1}(a+R^l+R^r)(\partial_+\kappa -[\kappa,R^l \partial_+\kappa]), \partial_-\kappa -[\kappa,R^l \partial_-\kappa]\biggr)_\K.\label{earb'}\ee 
It remains to determine the operators $(E^\na+M^R)\tilde\Pi^R(\tilde k_R)$ and $(E^{\ma}+M^L)\tilde\Pi^L(\tilde k_L)$, which are both the  endomorphisms of the vector space $\K$. We deduce from the formulae \eq{dspl} and \eq{39}, that
\be  \{\tilde f_1,\tilde f_2\}_{\tilde K_L}(\tilde k_L)=\left(\tilde\nabla^l \tilde f_1,T^i_L\right)_{\D_0}\left(t_i,\tilde\Pi^L(\tilde k_L)t_j\right)_{\D_0}\left(T^L_j,\tilde\nabla^l \tilde f_2\right)_{\D_0}=\left(t_i,\tilde\Pi^L(\tilde k_L)t_j\right)_{\D_0}  \tilde\nabla^l_{T^i_L} \tilde f_1 \tilde\nabla^l_{T^j_L} \tilde f_2;\label{10l}\ee
\be  \{\tilde f_1,\tilde f_2\}_{\tilde K_R}(\tilde k_R)=\left(\tilde\nabla^r \tilde f_1,T^i_R\right)_{\D_0}\left(t_i,\tilde\Pi^R(\tilde k_R)t_j\right)_{\D_0}\left(T^R_j,\tilde\nabla^l \tilde f_2\right)_{\D_0}= \left(t_i,\tilde\Pi^R(\tilde k_R)t_j\right)_{\D_0} \tilde\nabla^l_{T^i_R} \tilde f_1 \tilde\nabla^l_{T^j_R} \tilde f_2,\label{10r}\ee
where $t_i\in\K$  is the orthonormal basis
  on the compact simple Lie algebra $\K$ verifying 
\be (t_i,t_j)_\K=-\delta_{ij},\label{nr}\ee
 and the basis $T^i_L\in\tilde\K_L$ and $T^i_R\in\tilde\K_R$ were introduced in \eq{20} and in \eq{308}.
On the other hand, from Eqs.\eq{plr}, we infer 
\be \{\tilde f_1,\tilde f_2\}_{\tilde K_L}(\tilde k_L)=-\left(t_i,Ad_{\tilde k_L}t_m\right)_{\D_0}\left(Ad_{\tilde k_L}T_L^m,t_j\right)_{\D_0}\tilde\nabla^l_{T^i_L} \tilde f_1\tilde\nabla^l_{T^j_L} \tilde f_2;\label{11l}\ee
\be \{\tilde f_1,\tilde f_2\}_{\tilde K_R}(\tilde k_R)=-\left(t_i,Ad_{\tilde k_R}t_m\right)_{\D_0}\left(Ad_{\tilde k_R}T_R^m,t_j\right)_{\D_0}\tilde\nabla^l_{T^i_R} \tilde f_1\tilde\nabla^l_{T^j_R} \tilde f_2;\label{11r}\ee
Combining Eqs.\eq{10l}, \eq{10r} with  Eqs.\eq{11l}, \eq{11r}, we find
\be \left(t_i,\tilde\Pi^L(\tilde k_L)t_j\right)_{\D_0} =-\left(t_i,Ad_{\tilde k_L}t_m\right)_{\D_0}\left(Ad_{\tilde k_L}T_L^m,t_j\right)_{\D_0},\label{12l}\ee
\be \left(t_i,\tilde\Pi^R(\tilde k_R)t_j\right)_{\D_0}=-\left(t_i,Ad_{\tilde k_R}t_m\right)_{\D_0}\left(Ad_{\tilde k_R}T_R^m,t_j\right)_{\D_0}.\label{12r}\ee
We can rewrite the formulae \eq{12l}, \eq{12r} without resorting to the choices of the basis by writing
\be \tilde\Pi^L(\tilde k_L)=-\tilde\J_LAd_{\tilde k_L}\J_L Ad_{\tilde k^{-1}_L}\J_L;\label{13l}\ee
\be \tilde\Pi^R(\tilde k_R)=-\tilde\J_RAd_{\tilde k_R}\J_R Ad_{\tilde k^{-1}_R}\J_R.\label{13r}\ee
Here all the operators $\J_L,\J_R,\tilde \J_L,\tilde \J_R:\D_0\to\D_0$  are projectors; $\J_L$ projects on $\K$ with the kernel $\tilde\K_L$,
$\J_R$ projects on $\K$ with the kernel $\tilde\K_R$, $\tilde \J_L$ projects on $\tilde\K_L$ with the kernel $\K$ and
 $\tilde \J_R$ projects on $\tilde\K_R$ with the kernel $\K$. With the choice of the basis $t_i,T_L^i,T_R^i$ as before we have, in particular
 \be \J_L A=(A,T^i_L)_{\D_0}\  t_i,\quad  \J_R A=(A,T^i_R)_{\D_0} \ t_i, \quad \tilde \J_L A= (A,t_i)_{\D_0}\  T^i_L,\quad \tilde \J_R A= (A,t_i)_{\D_0} \ T^i_R,\quad A\in\D_0.\ee
Thus, for the subalgebras 
\be \tilde\K_L=\{(-R^l\kappa,\kappa),\kappa\in\K\},\quad \tilde \K_R=\{ (-R^r\kappa,\kappa),\kappa\in\K\}\label{dga}\ee
we find
 \be \J_L(x,y)=(x+R^ly,0), \J_R(x,y)=(x+R^ry,0),   \tilde \J_L (x,y)= (-R^ly,y), \tilde \J_R (x,y)= (-R^ry,y), (x,y)\in\D_0.\label{pro}\ee
Using the formulae \eq{ind}, \eq{adj}, \eq{dgs}, \eq{13l}, \eq{13r} and \eq{pro}, we find 
  \be \tilde\Pi^L(\kappa)x=-\left(-R^l\left([\kappa,x]-[\kappa,R^l[\kappa,x]]\right),[\kappa,x]-[\kappa,R^l[\kappa,x]] \right),\quad x\in\K;\ee
\be \tilde\Pi^R(\kappa)x=-\left(-R^r\left([\kappa,x]-[\kappa,R^r[\kappa,x]]\right),[\kappa,x]-[\kappa,R^r[\kappa,x]] \right),\quad x\in\K.\ee
Furthermore, from the formulae \eqref{ho} and  \eqref{mlr} we infer
 \be (E^\na+M^R)\tilde\Pi^R(\kappa)x=-(a+R^l+R^r)\left([\kappa,x]-[\kappa,R^r[\kappa,x]]\right);\ee
  \be (E^{\ma}+M^L)\tilde\Pi^L(\kappa)x= -(a+R^l+R^r)\left([\kappa,x]-[\kappa,R^l[\kappa,x]]\right).\ee
We can now finally rewrite the formulae \eq{ealb} and \eq{earb} in the final form
\be \tilde S_{\na}=\jp\int d\tau \oint \biggl(\Bigl((a+R^l+R^r)^{-1}-O^r(\kappa) {\rm ad}_\kappa\Bigl)^{-1}O^r(\kappa)\partial_+\kappa, O^r(\kappa)\partial_-\kappa  \biggr)_\K;\label{efia}\ee
 \be \tilde S_{\ma}=\jp\int d\tau \oint \biggl(\Bigl((a+R^l+R^r)^{-1}-O^l(\kappa) {\rm ad}_\kappa\Bigl)^{-1}O^l(\kappa)\partial_+\kappa, O^l(\kappa)\partial_-\kappa  \biggr)_\K,\label{efib}\ee
 where the operators $O^{l,r}(\kappa):\K\to\K$  are defined as
 \be O^l(\kappa):= 1 - {\rm ad}_\kappa\circ  R^l, \quad O^r(\kappa):= 1 - {\rm ad}_\kappa \circ R^r.\ee
 One of the important results of the present article is the statement that the $\sigma$-models \eq{efia} and\eq{efib} are dual to each other and the T-duality which relies them is not the Poisson-Lie one but the more general affine Poisson T-duality. Of course, if we want that this new duality be really interesting, we have to show that the $\sigma$-models \eq{efia} and \eq{efib} cannot be rendered equivalent by field redefinitions, or, equivalently, by a point canonical transformation.  To show that, 
 we start the argument by noting that when  $R^r=R^l=R$ (which corresponds to the special case when affine Poisson T-duality
 becomes the Poisson-Lie T-duality) then the $\sigma$-models \eq{efia} and\eq{efib} obviously coincide. The whole question
 is what happens if we get  out slightly  of this special case and
 consider a deformation  $R^l=R+\delta R$ and $R^r=R-\delta R$. Do the deformed models \eq{efia} and \eq{efib} remain the same up to  field redefinitions?  If yes, this would mean that there exists a vector field $\B$ on the target which would verify the following condition
 \be \L_{\B}\T^{+}=\T^+-\T^-,\label{652}\ee
 where $\T^\pm$  is defined as the following section of the tensor product of the tangent bundle of the target with itself:
 \be \T^\pm:=\biggl(\Bigl((a+2R)^{-1}-\left(1 - {\rm ad}_\kappa\circ (R\pm\delta R))\circ  {\rm ad}_\kappa \right)\Bigr) \ ^{R\pm\delta R}\tilde\nabla^l,\ ^{R\pm\delta R}\tilde\nabla^l\biggl)_\K.\label{655}\ee 
 Here the $\K$-valued vector fields $\ ^{R\pm\delta R}\tilde\nabla^l$ are dual to the right-invariant Maurer-Cartan forms $\left(1 - {\rm ad}_\kappa\circ  (R\pm\delta R)\right)d\kappa$. Explicitely, we have in the orthonormal basis $t_i$ (cf. Eq.\eq{nr}) 
 \be  \ ^{R\pm\delta R}\tilde\nabla^l \equiv t_i\ ^{R\pm\delta R}\tilde\nabla^l_{t_i},\label{656}\ee
 where 
  \be \ ^{R\pm\delta R}\tilde\nabla^l_{t_i}=\left(\delta_{pi}+\left([(R\pm\delta R)t_i,\kappa],t_p\right)_\K\right) \d_{\kappa_p}\label{658}\ee
  and   the coordinates $\kappa_p$ on the target space  are defined by the decomposition
  \be \kappa=\kappa_p t_p.\ee 
  Note that the geometric quantities $\T^\pm$ are dual with respect to those appearing in the Lagrangian and we have chosen them in order to get rid of the uncomfortable inverse.
  
  Let us rewrite the formulae \eq{655}, \eq{656} and \eq{658} more invariantly. For that,  we define a $\K$-valued partial derivative $\d_\kappa$ as
  \be \d_\kappa=t_p\d_{\kappa_p}.\ee
  Then 
  \be  \ ^{R\pm\delta R}\tilde\nabla^l=\d_\kappa+(R\pm\delta R){\rm ad}_\kappa\d_\kappa \label{666}\ee
  and
   \be \T^\pm:=\biggl( \Bigl((a+2R)^{-1}(1+(R\pm\delta R)\ad_\kappa)-\ad_\kappa\Bigr)\dk,\left(1+(R\pm\delta R)\ad_\kappa \right)\dk\biggr)_\K.
   \label{670}\ee
  We find also up to first order
  $$ \T^+-\T^-=$$\be =\biggl( \Bigl((a+2R)^{-1}(1+ R \ad_\kappa)-\ad_\kappa\Bigr)\dk, 2(\delta R)\ad_\kappa \dk\biggr)_\K+
  \biggl( \Bigl((a+2R)^{-1}2(\delta R)\ad_\kappa) \Bigr)\dk,\left(1+ R \ad_\kappa \right)\dk\biggr)_\K.
   \label{671}\ee 
  We shall now look for the vector field $\B$ satisfying the condition \eq{652} in the form
\be \B=(B(\kappa),\dk)_\K,\ee
where $B(\kappa)$ is a $\K$-valued function on the target.  By counting the powers of the variable $\kappa$ on both sides of Eq.\eq{652}, we find that $B(\kappa)$ must be at most linear in $\kappa$ and it must be also homogeneous because for $\kappa=0$ the quantities $\T^+$ and $\T^-$ coincide. This means
\be B(\kappa)=B\kappa,\ee
 where $B:\K\to\K$ is a linear operator. If we now work out the condition \eq{652} and set $\kappa=0$, we obtain
 \be \left((a+2R)^{-1}B^*\d_\kappa,\d_\kappa\right)_\K+ \left((a+2R)^{-1}\d_\kappa,B^*\d_\kappa\right)_\K=0,\label{679}
 \ee
 where $B^*$ is adjoint to $B$ with respect to the bilinear form $(.,.)_\K$. From Eq. \eq{679} we then deduce
 \be (a+2R)B+B^*(a+2R)=0,\label{683}\ee
  and, by taking the adjoint of this relation, also
   \be (a-2R)B+B^*(a-2R)=0.\label{684}\ee
 By adding as well as substracting Eqs. \eq{683} and \eq{684}, we infer that
 $B$ is anti-Hermitian and it commutes with $R$:
 \be B^*=-B,\quad [R,B]=0.\label{688}\ee
  We use the relation \eq{688} for writing the part of the condition \eq{652} linear in $\kappa$ as follows
 \be \biggl(\frac{a}{2a+4R}(\ad_\kappa)_B\d_\kappa+(\ad_\kappa)_B \frac{a}{2a+4R}\d_\kappa,\d_\kappa\biggr)_\K= \biggl(\frac{2}{a+2R}(\delta R)\ad_\kappa\d_\kappa+\ad_\kappa(\delta R)\frac{2}{a+2R}\d_\kappa,\d_\kappa\biggr)_\K,\label{690}\ee
 where
 \be (\ad_\kappa)_B\equiv \ad_\kappa B -B\ \!\ad_\kappa   +\ad_{(B\kappa)} .\ee
We now rewrite the condition \eq{690} as 
\be ({2a+4R})\Bigl(a(\ad_\kappa)_B -4(\delta R)\ad_\kappa\Bigr) +\Bigl(a(\ad_\kappa)_B-4\ad_\kappa(\delta R)\Bigr) ({2a+4R})=0\label{695}
 \ee
 and the conjugated one as
\be ({2a-4R})\Bigl(-a(\ad_\kappa)_B -4(\delta R)\ad_\kappa\Bigr) +\Bigl(-a(\ad_\kappa)_B-4\ad_\kappa(\delta R)\Bigr) ({2a-4R})=0,\label{698}
 \ee
Adding and subtracting \eq{695} and \eq{698}, we find
 \be -(\delta R)\ad_\kappa -\ad_\kappa(\delta R)+R(\ad_\kappa)_B+(\ad_\kappa)_B R=0\ee
\be a^2(\ad_\kappa)_B - 4 R(\delta R)\ad_\kappa -4\ad_\kappa(\delta R)R=0.\ee
Inserting the second relation into the first one we obtain
\be  -a^2(\delta R)\ad_\kappa -a^2\ad_\kappa(\delta R) +4R(\delta R)\ad_{\kappa}R+4R^2(\delta R)\ad_{\kappa}+4R \ad_{\kappa}(\delta R)R+4\ad_{\kappa}(\delta R)R^2=0.\ee
This relation must be true for all $a$ which means that $\delta R$ must anticommute with the adjoint action of every element of $\K$ which is impossible.
 This implies that the vector field $\B$ does not exist and the actions $\tilde S_\na$ and $\tilde S_{\ma}$ given by Eqs. \eq{efia} and \eq{efib} cannot be made equivalent by field redefinitions.

\section{Affine quasi-Poisson groups} 
 \setcounter{equation}{0}

Recall from Ref. \cite{AK} that the quasi-Poisson $\K$-space is a manifold $(M,\pi)$ on which acts a Lie quasi-bialgebra $\K$ in a way compatible
with the quasi-Poisson bivector $\pi$. Recall that the Lie quasi-bialgebra $\K$ is an ordinary Lie algebra $(\K,[.,.])$ supplied with additional anti-symmetric bracket $[.,.]^*$ on the dual space $\K^*$ and equipped also with  a completely antisymmetric trilinear form $\chi:\Lambda^3\K^*\to\br$. The structures $[.,.]$,
$[.,.]^*$ and $\chi$  must be compatible in the sense that the
direct sum of the vector spaces $\D^q:=\K\oplus\K^*$ has to be ordinary Lie algebra equipped with  the following commutator
\be [x\oplus\alpha, y\oplus \beta]_{\D^q}=\left([x,y]+x\circ[\beta,.]^*-y\circ[\alpha,.]^*+\chi(\alpha,\beta,.)\right)\oplus\left([\alpha,\beta]^*-\beta\circ [x,.]+\alpha\circ [y,.]\right).\label{qdc}\ee
Here e.g. the expression  $x\circ[\beta,.]^*$ has to be interpreted as the element of $\K$ which acts on the elements of $\K^*$ as
\be \la x\circ[\beta,.]^*,\gamma\ra:=\la x,[\beta,\gamma]^*\ra,\quad \gamma\in\K^*.\ee
We notice, that if the trilinear form $\chi$ vanishes then the Lie quasi-bialgebra $\K$ is just the standard Lie bialgebra, that is  the bracket 
$[.,.]^*$ gives the Lie commutator on the dual space $\K^*$ and the commutator \eq{qdc} is the one of the standard Drinfeld double $\D$ of $\K$. If the quantity $\chi$ does not vanish, the anti-symmetric  bracket $[.,.]^*$ may be but need not be a Lie commutator on $\K^*$, nevertheless the bracket \eq{qdc} on $\D^q:=\K\oplus\K^*$ is always an honest Lie commutator.
We shall refer to $\D^q$ as to the quasi-Drinfeld double of the Lie quasi-bialgebra. 

We shall see concrete examples of the Lie quasi-bialgebras in the next section, here we continue for the moment the general exposure. An action $\rho$ of the Lie quasi-bialgebra $\K$ on the manifold $(M,\pi)$ is called quasi-Poisson, if it holds
\be \L_{\rho(x)}\pi=-\rho(\tilde f(x));\label{qp1} \ee
\be \jp[\pi,\pi]_M=\rho(\chi).\label{qp2} \ee
Here $\L_{\rho(x)}$ stands for the Lie derivative, $\tilde f:\K\to\K\wedge\K$ is the map dual to the bracket $[.,.]^*:\K^*\wedge\K^*\to\K^*$ and $[.,.]_M$ is the Schouten bracket on the  manifold $M$. We recall that, for decomposable bivectors, the Schouten bracket is defined as
\be [u_1\wedge u_2,v_1\wedge v_2]_M:=[u_1,v_1]_M\wedge u_2\wedge v_2+u_1\wedge [u_2,v_1]_M\wedge v_2- v_1\wedge [u_1,v_2]_M\wedge u_2-v_1\wedge u_1\wedge[u_2,v_2]\label{bsc}\ee
and the general bracket is obtained by linearity. Of course, $[u_1,v_1]_M$ stands for the Lie bracket of vector fields. 

It may be illuminating to write the conditions \eq{qp1}, \eq{qp2} for the quasi-Poisson action in some basis $t_i$ of the Lie quasi-bialgebra $\K$. We have
\be  \L_{\rho(t_i)}\pi=-\jp\tilde f_i^{\ jk}\rho(t_j)\wedge \rho(t_k);\label{qp3} \ee
\be \jp[\pi,\pi]_M=\frac{1}{6}\chi^{ijk}\rho(t_i)\wedge \rho(t_j)\wedge \rho(t_k).\label{qp4} \ee

Let us now define the affine quasi-Poisson group $K$ as the Lie group equipped with a bivector $\pi^\qa$, on which the standard left and right group  multiplications induce the quasi-Poisson actions (possibly with respect to two non-isomorphic Lie quasi-bialgebras) in the sense of the conditions \eq{qp1} and \eq{qp2}. The left
action  $\rho_L$ and the right one $\rho_R$ are obviously described by the invariant vector fields 
\be \rho_L(t_i)=\nabla^l_{t_i},\quad  \rho_R(t_i)=-\nabla^r_{t_i},\label{ga7}\ee
so that we require in full analogy with \eq{pma} and \eq{pmb}
\be \L_{\nabla^l_{t_i}}\pi^\qa= -\jp \ ^{L}\tilde c_i^{\ jk}\nabla^l_{t_j}\wedge \nabla^l_{t_k};\label{pmc}\ee
\be \L_{\nabla^r_{t_i}}\pi^\qa= \jp\ ^{R}\tilde c_i^{\ jk}\nabla^r_{t_j}\wedge \nabla^r_{t_k}.\label{pmd}\ee
Here $t_i$ is some basis of the Lie algebra $\K$ of the group $K$, $\ ^{L}\tilde c_i^{\ jk}$ and $\ ^{R}\tilde c_i^{\ jk}$
are, respectively, the structure constants of the brackets $[.,.]^*_L$ and  $[.,.]^*_R$. The condition \eq{qp2} in this context becomes 
\be \jp[\pi^\qa,\pi^\qa]_K= -\frac{1}{6}\chi_R^{ijk}\nabla^r_{t_i}\wedge \nabla^r_{t_k}\wedge \nabla^r_{t_l}=\frac{1}{6}\chi_L^{ijk}\nabla^l_{t_i}\wedge \nabla^l_{t_j}\wedge \nabla^l_{t_k}.\label{qp6}\ee

It is clear that our definition of the affine quasi-Poisson group implies that the Lie quasi-bialgebra $\K_L:=(K,[.,.],[.,.]_L^*,\chi_L)$ acting from the left is not necessary isomorphic to the  Lie quasi-bialgebra $\K_R:=(K,[.,.],[.,.]_R^*,\chi_R)$ acting for the right, however, the fact that the both left and right actions are simultaneously quasi-Poisson with respect to the same bivector $\pi^\qa$  implies some restrictions on the possibles structures of the Lie quasi-bialgebras 
$\K_L$ and $\K_R$. In particular, it must hold   \be \chi_L=-\chi_R\equiv \chi,\ee and, moreover, $\chi$ must be ad$_\K$-invariant.  This is not all, however. It turns out also that the dual brackets 
$[.,.]_L^*$ and $[.,.]_R^*$ must be both Lie commutators, i.e. the Jacobi identity must hold for them. To see this, we calculate the Lie derivatives of Eq. \eq{qp6} with respect to the left-invariant and right-invariant vector fields $\nabla^r_{t_i}$ and $\nabla^l_{t_i}$. Because of the invariance of the trilinear form $\chi$, the Lie derivatives of the right-hand-side vanish, while for the Lie derivatives of the left-hand-side, we obtain successively from Eqs. \eq{pmc}, \eq{pmd} and \eq{bsc}
$$  \L_{\nabla^l_{t_i}}[\pi^\qa,\pi^\qa]_K=2[\L_{\nabla^l_{t_i}}\pi^\qa, 
\pi^\qa]= -\ ^{L}\tilde c_i^{\ jk}[\nabla^l_{t_j}\wedge \nabla^l_{t_k},\pi^\qa]= -2\ ^{L}\tilde c_i^{\ jk}\nabla^l_{t_j}\wedge [\nabla^l_{t_k},  \pi^\qa]_K=$$
\be \ ^{L}\tilde c_i^{\ jk}\ ^{L}\tilde c_k^{\ mn}\nabla^l_{t_j}\wedge \nabla^l_{t_m}\wedge \nabla^l_{t_n}=\frac{1}{3}\left(\ ^{L}\tilde c_i^{\ jk}\ ^{L}\tilde c_k^{\ mn}+\ ^{L}\tilde c_i^{\ mk}\ ^{L}\tilde c_k^{\ nj}+\ ^{L}\tilde c_i^{\ nk}\ ^{L}\tilde c_k^{\ jm}\right)\nabla^l_{t_j}\wedge \nabla^l_{t_m}\wedge \nabla^l_{t_n}=0.\label{ji}\ee 
Thus we infer 
\be \ ^{L}\tilde c_i^{\ jk}\ ^{L}\tilde c_k^{\ mn}+\ ^{L}\tilde c_i^{\ mk}\ ^{L}\tilde c_k^{\ nj}+\ ^{L}\tilde c_i^{\ nk}\ ^{L}\tilde c_k^{\ jm}=0,\ee
which is nothing but the Jacobi identity for the structure constants $\ ^{L}\tilde c_i^{\ jk}
$. The  Jacobi identity for the structure constants $\ ^{R}\tilde c_i^{\ jk}$
can be established similarly.

 As in the case of  the theory of the affine Poisson groups treated in Section 3, also in the affine quasi-Poisson case the crucial role is played by the value  of the bivector field $\pi^\qa$ at the unit element $e_K$ of the group $K$.
We  again denote this value as $m$:
\be m:=\pi^\qa(e_K).\ee
 Our next goal is to show, that there are two Poisson-Lie structures $\pi^L$ and $\pi^R$ naturally associated with the affine quasi-Poisson structure $\pi^\qa$. They are given by the formulae
 \be \pi^L=L_*m-\pi^\qa,\quad \pi^R=\pi^\qa-R_*m.\label{717}\ee
 Let us prove  e.g.  that $\pi^L$ is the Poisson-Lie structure. For that, we have to prove the following three identities:
 \be [\pi^L,\pi^L]_K=0;\label{1d}\ee
\be \L_{\nabla^l_{t_i}}\pi^L= \jp \ ^{L}\tilde c_i^{\ jk}\nabla^l_{t_j}\wedge \nabla^l_{t_k};\label{2d}\ee
\be \L_{\nabla^r_{t_i}}\pi^L= \jp\ ^{L}\tilde c_i^{\ jk}\nabla^r_{t_j}\wedge \nabla^r_{t_k}.\label{3d}\ee
The condition which is the easiest to prove is Eq.\eq{2d}, because $\nabla^r_{t_i}$ commutes with $\nabla^l_{t_j}$, hence 
\be \L_{\nabla^l_{t_i}}\pi^L=\L_{\nabla^l_{t_i}}\left(\jp m^{jk}
\nabla^r_{t_j}\wedge \nabla^r_{t_k}-\pi^\qa\right)=-\L_{\nabla^l_{t_i}}\pi^\qa =\jp\ ^{L}\tilde c_i^{\ jk}\nabla^l_{t_j}\wedge \nabla^l_{t_k},\label{aa}\ee
where in the last equality we have used Eq.\eq{pmc}.

Now we prove Eq. \eq{3d}. We first note that the bivector $\pi_L$ vanishes at the group origin $e_K$, therefore the Lie derivatives $\L_{\nabla^l_{t_i}}\pi^L$ and $\L_{\nabla^r_{t_i}}\pi^L$ coincide at $e_K$. This means that in order to prove Eq.\eq{3d}, it is sufficient  to prove that $\L_{\nabla^r_{t_i}}\pi^L$ is a left-invariant bivector, because such bivectors are completely determined by their values at the group origin. The proof of the left invariance of $\L_{\nabla^r_{t_i}}\pi^L$ is simple, indeed, we 
find from Eq.\eq{pmd}
\be \L_{\nabla^r_{t_i}}\pi^L=\L_{\nabla^r_{t_i}}\left(\jp m^{jk}\nabla^r_{t_j}\wedge \nabla^r_{t_k}-\pi^\qa\right)=\left(-\jp\ ^{R}\tilde c_i^{\ lk} +\jp f_{ij}^{\ \ l}m^{jk}-\jp f_{ij}^{\ \ k}m^{jl}  \right)\nabla^r_{t_l}\wedge \nabla^r_{t_k}.\label{bb}\ee
As a bonus, we have obtained the following identity relying the left and the right dual structure constants
\be \ ^{L}\tilde c_i^{\ lk}=-\ ^{R}\tilde c_i^{\ lk} +f_{ij}^{\ \ l}m^{jk}-f_{ij}^{\ \ k}m^{jl},\ee
where $f_{ij}^{\ \ k}$ are the structure constants of the Lie algebra $\K$.

In ordre to prove \eq{1d}, we need first to establish the following relation
\be \left(-\ ^{L}\tilde c_j^{\ kn}+\ ^{R}\tilde c_j^{\ kn}\right)m^{ij}+
\left(-\ ^{L}\tilde c_j^{\ ni}+\ ^{R}\tilde c_j^{\ ni}\right)m^{kj}+
\left(-\ ^{L}\tilde c_j^{\ ik}+\ ^{R}\tilde c_j^{\ ik}\right)m^{nj}=2\chi^{kni}.\label{fre}\ee
To do it, let us calculate the value of the Schouten bracket $[\pi^\qa,\pi^\qa]_K$ et the unit element $e_K$. For that, we represent the bivector $\pi^\qa$ in terms of the left and the right trivialisation of the bundle $\Lambda^2TK$ as
\be \pi^\qa\equiv\jp\ ^r\Pi^{ij}\nabla^r_{t_i}\wedge\nabla^r_{t_j}\equiv\jp\ ^l\Pi^{kn}\nabla^l_{t_k}\wedge\nabla^l_{t_n}.\ee
Because $\nabla^r_{t_i}$ commutes with $\nabla^l_{t_j}$, we find
\be [\pi^\qa,\pi^\qa]_K=\jp\ ^r\Pi^{ij}\left(\nabla^r_{t_j}\ ^l\Pi^{kn}\right)\nabla^r_{t_i}\wedge\nabla^l_{t_k}\wedge\nabla^l_{t_n}+
\jp\ ^l\Pi^{ij}\left(\nabla^l_{t_j}\ ^r\Pi^{kn}\right)\nabla^l_{t_i}\wedge\nabla^r_{t_k}\wedge\nabla^r_{t_n},\label{bl4}\ee
which, 
with the help of Eqs. \eq{pmc} and \eq{pmd}, gives at the group unit $e_K$
\be [\pi^\qa,\pi^\qa]_K\biggl\vert_{e_K}=\frac{1}{2}\left(-\ ^{L}\tilde c_j^{\ kn}+\ ^{R}\tilde c_j^{\ kn}\right)m^{ij}t_i\wedge t_k \wedge t_n.\ee
From this and from Eq.\eq{qp6}, we infer finally the desired identity \eq{fre}.

We now have from Eqs. \eq{2d} and \eq{3d}
\be [\nabla^r_{t_i},L_*m]_K=[\nabla^r_{t_i},\pi^\qa]_K+[\nabla^r_{t_i},\pi^L ]_K=\jp\left(\ ^{R}\tilde c_i^{\ kn}+\ ^{L}\tilde c_i^{\ kn}\right)\nabla^r_{t_k}\wedge\nabla^r_{t_n},  \ee
hence we find 
\be [L_*m,L_*m]_K=\jp\left[m^{ij}\nabla^r_{t_i}\wedge \nabla^r_{t_j},L_*m\right]_K=\jp m^{ij}\left(\ ^{R}\tilde c_j^{\ kn}+\ ^{L}\tilde c_j^{\ kn}\right) \nabla^r_{t_i}\wedge\nabla^r_{t_k}\wedge\nabla^r_{t_n}.\label{cc}\ee
Then we have from \eq{pmd} 
\be [\pi^\qa,L_*m]_K=\jp\left[\pi^\qa,m^{ij}\nabla^r_{t_i}\wedge \nabla^r_{t_j}\right]_K=\jp m^{ij} \ ^{R}\tilde c_j^{\ kn}\nabla^r_{t_i}\wedge\nabla^r_{t_k}\wedge\nabla^r_{t_n}.\label{ff}\ee
By using Eqs.\eq{qp6}, \eq{cc}, \eq{ff} and \eq{fre}, we find finally
\be [\pi^L,\pi^L]_K=[L_*m-\pi^\qa,L_*m-\pi^\qa]_K=\left(\frac{1}{3}\chi^{ikn}- \jp m^{ij}\left(-\ ^{L}\tilde c_j^{\ kn}+\ ^{R}\tilde c_j^{\ kn}\right) \right)  \nabla^r_{t_i}\wedge \nabla^r_{t_k}\wedge \nabla^r_{t_n}=0. \label{dos}\ee
The bivector $\pi^L\equiv L_*m-\pi^\qa$ thus defines the Poisson-Lie structure on the group $K$ and it can be shown similarly that the bivector
$\pi^R\equiv\pi^\qa-R_*m$ also defines the Poisson-Lie structure. 

Given the affine quasi-Poisson structure $\pi^\qa$, does it exist the mirror affine quasi-Poisson structure $\pi^\aq$ similarly as it is the case in the ordinary affine Poisson case? We now show that the answer to this question is affirmative. We define this mirror affine quasi-Poisson structure  by   the formula  
\be \pi^{\aq}:= R_*m+L_*m-\pi^\qa,\label{760}\ee
in full analogy with Eq.\eq{263}. Of course,
we must show that $\pi^{\aq}$ indeed verifies the affine quasi-Poisson conditions  \eq{pmc}, \eq{pmd}
and \eq{qp6} with the  role of the left and the right inversed. In other words, we have to prove that
\be \L_{\nabla^l_{t_i}}\pi^\aq= - \jp\ ^{R}\tilde c_i^{\ jk}\nabla^l_{t_j}\wedge \nabla^l_{t_k};\label{764}\ee
\be \L_{\nabla^r_{t_i}}\pi^\aq= \jp\ ^{L}\tilde c_i^{\ jk}\nabla^r_{t_j}\wedge \nabla^r_{t_k};\label{765}\ee
\be \jp[\pi^\aq,\pi^\aq]_K=-\frac{1}{6} \chi^{ijk}\nabla^l_{t_i}\wedge \nabla^l_{t_j}\wedge \nabla^l_{t_k}=-\frac{1}{6}\chi^{ijk}\nabla^r_{t_i}\wedge \nabla^r_{t_j}\wedge \nabla^r_{t_k}.\label{762}\ee
 We start by proving \eq{765}
  \be \L_{\nabla^r_{t_i}}\pi^\aq=\L_{\nabla^r_{t_i}}(\pi^L+R_*m)=\L_{\nabla^r_{t_i}}\pi^L=\jp\ ^{L}\tilde c_i^{\ jk}\nabla^r_{t_j}\wedge \nabla^r_{t_k},\ee
  where we have used Eq. \eq{3d} as well as the fact that the right-invariant vector fields commute with the left-invariant ones.
 We prove \eq{764} similarly
 \be \L_{\nabla^l_{t_i}}\pi^\aq=\L_{\nabla^l_{t_i}}(L_*m-\pi^R)=-\L_{\nabla^l_{t_i}}\pi^R=-\jp\ ^{R}\tilde c_i^{\ jk}\nabla^l_{t_j}\wedge \nabla^l_{t_k}.\ee
 Finally, we establish the identity \eq{762} by using Eqs. \eq{cc}, \eq{ff} and \eq{fre}
$$[\pi^\aq,\pi^\aq]_K=\left[L_*m-\pi^R,L_*m-\pi^R\right]_K=[\pi^R,\pi^R]_K+
 [L_*m,L_*m]_K-2[L_*m,\pi^R]=$$
 \be = [L_*m,L_*m]_K-2[L_*m,\pi^\qa]_K=  - \jp m^{ij}\left(-\ ^{L}\tilde c_j^{\ kn}+\ ^{R}\tilde c_j^{\ kn}\right)   \nabla^r_{t_i}\wedge \nabla^r_{t_k}\wedge \nabla^r_{t_n}=-\frac{1}{3}
 \chi^{ikn} \nabla^r_{t_i}\wedge \nabla^r_{t_k}\wedge \nabla^r_{t_n}.\ee

It thus turns out that we can associate to every affine quasi-Poisson group $(K,\pi^\qa)$ two Drinfeld doubles\footnote{It has to be stressed that the Drinfeld doubles $\D_L$ and $\D_R$ need not to be respectively isomorphic to the quasi-Drinfeld doubles $\D^q_L$ and $\D^q_R$ of the Lie quasi-bialgebras $\K_L$ and $\K_R$.} $D_L$ and $D_R$, which are, respectively, the Drinfeld doubles  of the Poisson-Lie groups $(K,\pi^L)$ and $(K,\pi^R)$.
The double $D_L$ has naturally two maximally isotropic subgroups $K$ and $\tilde K_L$ , while $D_R$ has subgroups $K$ and $\tilde K_R$. We call the Lie groups $\tilde K_L$ and $\tilde K_R$  the left and the right dual groups of the affine quasi-Poisson group $K$;  their Lie algebras are isomorphic
to the Lie algebras $\tilde\K_L$ and $\tilde\K_R$ with the structure constants $\ ^{L}\tilde c_i^{\ jk}$ and $\ ^{R}\tilde c_i^{\ jk}$. 

For completeness, we detail how the left and right Poisson-Lie structures $\pi^L$ and $\pi^R$ as well as the mirror affine quasi-Poisson structure $\pi^{\aq}$ associated to the affine Poisson structure $\pi^\qa$ are 
  extracted from the structure of the Drinfeld doubles $D_L$ and $D_R$ of the affine quasi-Poisson group $K$. We start from the perspective of the double $D_L$,  
  we pick a basis $t_i, i=1,...,d$ of the Lie algebra $\K\subset\D_L$ and  the dual  basis $T^i_L, i=1,...,d$ of $\tilde\K_L\subset\D_L$   in such a way that 
\be (t_i,T^j_L)_{\D_L}=\delta_i^j\label{834}\ee
We can then write
\be \ ^{L}\tilde c_i^{\ jk}=\left([T_L^j,T_L^k],t_i\right)_{\D_L}. \ee
The affine  quasi-Poisson bracket $\{.,.\}^{\qa}_K$ corresponding   to the affine quasi-Poisson bivector $\pi^{\qa}$ is then given by the formula
\be \{f_1,f_2\}_K^\qa(k)=\jp m^{ij}\nabla^r_{t_i}f_1\nabla^r_{t_j}f_2+(T^i_L,Ad_kT^j_L)_{\D_L}\nabla^l_{t_i}f_1\nabla^r_{t_j}f_2;\label{pba'}
\ee
and the Poisson-Lie bracket $\{.,.\}^{L}_K$ corresponding to the Poisson-Lie bivector $\pi^{L}$ is given by the formulae
\be \{f_1,f_2\}_K^L(k)=-(T^i_L,Ad_kT^j_L)_{\D_L}\nabla^l_{t_i}f_1\nabla^r_{t_j}f_2.\label{plr'}\ee
Here $f_1,f_2$ are smooth functions on $K$ and the Einstein summation convention applies.

From the point of view of the double $D_R$,  we pick a basis $t_i, i=1,...,d$ of the Lie algebra $\K\subset\D_R$ and the dual basis $T^i_R, i=1,...,d$ of $\tilde\K_R\subset\D_R$ in such a way that 
\be (t_i,T^j_R)_{\D_R}=\delta_i^j.\label{845}\ee
We can then write
\be \ ^{R}\tilde c_i^{\ jk}=\left([T_R^j,T_R^k],t_i\right)_{\D_R}.\ee
The mirror affine  quasi-Poisson bracket $\{.,.\}^{\aq}_K$ corresponding  to the mirror affine Poisson bivector $\pi^{\aq}$ is  then given by the formula
\be \{f_1,f_2\}_K^{\aq}(k)=\jp m^{ij}\nabla^r_{t_i}f_1\nabla^r_{t_j}f_2+(T^i_R,Ad_kT^j_R)_{\D_R}\nabla^l_{t_i}f_1\nabla^r_{t_j}f_2\label{pbam'}
\ee
and the Poisson bracket $\{.,.\}^{R}_K$ corresponding to the Poisson-Lie bivector $\pi^{R}$ is  given by  
\be   \{f_1,f_2\}_K^R(k)=-(T^i_R,Ad_kT^j_R)_{\D_R}\nabla^l_{t_i}f_1\nabla^r_{t_j}f_2.\label{314'}\ee
 
 \medskip
 
 \noindent {\bf Remark:} {The affine quasi-Poisson formulae \eq{pba'} and \eq{pbam'} look identical than the affine Poisson formulae \eq{pba} and \eq{pbam}, however, they are different in the sense that the matrix $m^{ij}$ has to be chosen in such a way that the brackets $\{.,.\}^\aq$ and
 $\{.,.\}^\na$
 be   quasi-Poisson in the first case and   Poisson in the second.}
 
 \medskip
 
 We finish this section by stating that the following useful relation holds  
 \be \{f_1\circ S,f_2\circ S\}_K^\qa=\{f_1,f_2\}_K^{\aq}\circ S,\label{876}\ee
 where $S:K\to K$ is defined as $S(k):=k^{-1}$. The proof of this fact is similar to  the one presented at the end of Section 3 in the affine Poisson case .

\section{Affine quasi-Poisson T-duality}
 \setcounter{equation}{0}
\subsection{General framework}
This section constitutes the generalization of Section 4 which was devoted to the 
affine Poisson T-duality.  

Let $(K,\pi^\qa)$  be the  affine quasi-Poisson group, $D_L$ and $D_R$ its Drinfeld doubles and $\tilde K_L\subset D_L$ and $\tilde K_R\subset D_R$ the dual groups of $K$ in the sense of Section 5.    Consider then  two $\sigma$-models living on $K$
\be S_\qa(k)=\jp\int d\tau \oint \biggl(\Bigl(E^\qa+\Pi^\qa(k)\Bigr)^{-1}\partial_+k k^{-1}, \partial_- k k^{-1}\biggr)_{\D_R};\label{817}\ee
\be S_{\aq}(k)=\jp\int d\tau \oint \biggl(\Bigl(E^{\aq}+\Pi^{\aq}(k)\Bigr)^{-1}\partial_+k k^{-1}, \partial_- k k^{-1}\biggr)_{\D_L},\label{818}\ee
 one $\sigma$-model on $\tilde K_R$
\be \tilde S_{\qa}(\tilde k_R)=\jp\int d\tau \oint \biggl(\Bigl(\tilde E^\qa+\tilde\Pi^R(\tilde k_R)\Bigr)^{-1}\partial_+\tilde k_R\tilde k_R^{-1}, \partial_-\tilde k_R\tilde k_R^{-1}\biggr)_{\D_R},\label{820}\ee
and, finally, one on $\tilde K_L$
 \be \tilde S_{\aq}(\tilde k_L)=\jp\int d\tau \oint \biggl(\Bigl(\tilde E^{\aq}+\tilde\Pi^L(\tilde k_L)\Bigr)^{-1}\partial_+\tilde k_L\tilde k_L^{-1}, \partial_-\tilde k_L\tilde k_L^{-1}\biggr)_{\D_L}.\label{822}\ee
 Here the fields-valued operators  $\Pi^\qa(k):\tilde\K_R\to\K$, $\Pi^\aq (k):\tilde\K_L\to\K$,   $\tilde\Pi^R(\tilde k_R):\K\to\tilde\K_R$ and $\tilde\Pi^L(\tilde k_L):\K\to\tilde\K_L$  characterize the affine quasi-Poisson structures $\pi^\qa$, $\pi^\aq$ on $K$ and the Poisson-Lie structures $\tilde\pi^L$,$\tilde \pi^R$ on $\tilde K_L$ and $\tilde K_R$ dual to the Poisson-Lie structures $\pi^L$,$\pi^R$ on  $K$ in the sense of the relations: 
 \be \la \pi^\qa,df_1\otimes df_2\ra(k)=\{f_1,f_2\}^\qa_K(k)=\left(\ \!^R\nabla^l f_1,\Pi^\qa(k)\ \!^R\nabla^l f_2\right)_{\D_R},\label{821}\ee
\be  \la \pi^\aq,df_1\otimes df_2\ra(k)=\{f_1,f_2\}^\aq_K(k)=\left(\ \!^L\nabla^l f_1,\Pi^{\aq}( k)\ \!^L\nabla^l f_2\right)_{\D_L},\label{822b}\ee
 \be  \la \tilde\pi^R,d\tilde f_1\otimes d\tilde f_2\ra(\tilde k_R) =\{\tilde f_1,\tilde f_2\}_{\tilde K_R}(\tilde k_R)=\left(\ \!^R\tilde\nabla^l \tilde f_1,\tilde\Pi^R(\tilde k_R)\ \!^R\tilde\nabla^l \tilde f_2\right)_{\D_R}.\label{827}\ee 
 \be  \la \tilde\pi^L,d\tilde f_1\otimes d\tilde f_2\ra(\tilde k_L) =\{\tilde f_1,\tilde f_2\}_{\tilde K_L}(\tilde k_L)=\left(\ \!^L\tilde\nabla^l \tilde f_1,\tilde\Pi^L(\tilde k_L)\ \!^L\tilde\nabla^l \tilde f_2\right)_{\D_L},\label{826}\ee

 Recall in this respect that $\ \!^L\nabla^r$ and $\ \!^R\nabla^r$ are, respectively, $\tilde\K_L$ and $\tilde\K_R$-valued differential operators acting on functions on $K$  as
\be \left(\ \!^{L}\nabla^l f, x\right)_{D_L}:= \nabla^l_{x} f, \qquad \left(\ \!^{R}\nabla^l f, x\right)_{D_R}:= \nabla^l_{x} f, \quad x\in \K,\label{824}\ee
and  $\ \!^{L}\tilde\nabla^{r}$, $\ \!^{R}\tilde\nabla^{r}$ are $\K$-valued differential operators acting respectively on the functions on $\tilde K_L$ and $\tilde K_R$ as
\be \left(\!^L\tilde\nabla^l \tilde f,\tilde x_L\right)_{\D_L}(\tilde k_L):= (\tilde\nabla^l_{\tilde x_L}\tilde f)(\tilde k_L)= \frac{d\tilde f(e^{s\tilde x_L}\tilde k_L)}{ds}\bigg\vert_{s=0}, \quad   \tilde x_L\in \tilde\K_L;\label{831}\ee
\be \left(\!^R\tilde\nabla^l \tilde f,\tilde x_R\right)_{\D_R}(\tilde k_R):= (\tilde\nabla^l_{\tilde x_R}\tilde f)(\tilde k_R)= \frac{d\tilde f(e^{s\tilde x_R}\tilde k_R)}{ds}\bigg\vert_{s=0}, \qquad   \tilde x_R\in \tilde\K_R.\label{832}\ee

  As in the affine Poisson context  studied in Section 4, also in the case of the affine quasi-Poisson T-duality the idea is to find suitable invertible linear operators 
  $E^\qa$, $E^\aq$, $\tilde E^\qa$ and $\tilde E^{\aq}$ in such a way  that the four $\sigma$-models 
  \eq{817}, \eq{818}, \eq{820} and \eq{822} are pairwise T-dual to each other. 
  
  In fact, the liberty of choice exists  only for the operator  $E^\qa:\tilde\K_R\to\K$ because the operators  $E^\aq:\tilde\K_L\to\K$, $\tilde E^\qa:\K\to\tilde\K_R$ and $\tilde E^{\aq}:\K\to\tilde\K_L$ turn out to be  determined from it. For the case of $\tilde E^\qa$ and $\tilde E^{\aq}$, this can be seen by rewriting the $\sigma$-model actions 
   \eq{817} and  \eq{818} equivalently as
 \be S_\qa(k)=\jp\int d\tau \oint \biggl(\Bigl(E^\qa+M^R+\Pi^R(k)\Bigr)^{-1}\partial_+k k^{-1}, \partial_- k k^{-1}\biggr)_{\D_R},\label{841}\ee
\be S_{\aq}(k)=\jp\int d\tau \oint \biggl(\Bigl(E^{\aq}+M^L+\Pi^L(k)\Bigr)^{-1}\partial_+k k^{-1}, \partial_- k k^{-1}\biggr)_{\D_L},\label{842}\ee
where 
 \be M^R=\Pi^\qa(e_K), \quad  M^L=\Pi^{\aq}(e_K)\label{843}\ee
 and $\Pi^L(k):\tilde\K_L\to\K$ and $\Pi^R(k):\tilde\K_R\to\K$ stand, respectively, for the left and right Poisson-Lie structures on $K$ (associated to the affine quasi-Poisson structure) in the sense of the formulae
\be \la \pi^L,df_1\otimes df_2\ra(k) =\{f_1,f_2\}^L_K(k)=\left(\ \!^L\nabla^l f_1,\Pi^L(k)\ \!^L\nabla^lf_2\right)_{\D_L},\label{844}\ee
 \be \la \pi^R,df_1\otimes df_2\ra(k) =\{f_1,f_2\}^R_K(k)=\left(\ \!^R\nabla^l f_1,\Pi^R(k)\ \!^R\nabla^lf_2\right)_{\D_R}.\label{846}\ee
We then observe that the $\sigma$-model \eq{841} is related by the standard Poisson-Lie T-duality to the model \eq{820} if the operator $\tilde E^\qa$ is inverse to
 $E^\qa+M_R$; and \eq{842} is Poisson-Lie T-dual to \eq{822} if  $\tilde E^\aq$ is inverse to $E^{\aq}+M_L$.
 
 We thus remark that if the $\sigma$-model \eq{817}
 is T-dual to \eq{818} then all four $\sigma$-models  \eq{817}, \eq{818}, \eq{820} and \eq{822} are pairwise T-dual to each other. From the point of view of the scheme \eqref{Sch2},   on the up-right  vertex of the scheme there is the $\sigma$-model \eqref{822}, on the up-left  there is  \eqref{818}, on the down-left  is the model \eqref{817}   and, finally, on the down-right  there is the $\sigma$-model \eqref{820}. The up-right vertex \eqref{822} is linked with the up-left vertex \eqref{818} by the standard Poisson-Lie T-duality based on the 
 subspace $\E_L=\{\tilde x +(E^\aq+M^L)\tilde x,\tilde x\in\tilde\K_L\}$ and the down-right vertex \eqref{820} is linked with the down-left vertex \eqref{817} by the standard Poisson-Lie T-duality based on the 
 subspace $\E_R=\{\tilde x +(E^{\qa}+M^R)\tilde x,\tilde x\in\tilde\K_R\}$, which means that the existence of  all horizontal arrow symplectomorphisms is established.

 \subsection{Sufficient condition for the affine quasi-Poisson T-duality}
 
For  what choice of the data $K,\Pi^\na,E^\qa,E^{\aq}$ the $\sigma$-model \eq{817}
 is T-dual to \eq{818}? In order to formulate the sufficient condition for this T-duality   we need to define two bilinear forms $(.,.)_\qa$, $(.,.)_{\aq}$  on the Lie algebra $\K$:
 \be (x,y)_\qa:=(x,(E^\qa)^{-1}y)_{\D_R},\quad (x,y)_{\aq}:=(x,(E^{\aq})^{-1}y)_{\D_L},\quad x,y\in \K.\label{867}\ee
 We then have  the following assertion
 
 \medskip 
 
 \noindent {\bf Assertion 3}: {\it If the bilinear form} $(.,.)_\qa$ {\it is symmetric, non-degenerate and 
 ad$_\K$-invariant and, moreover, if it coincides with the bilinear form} $(.,.)_{\aq}$ {\it then the 
 $\sigma$-model \eqref{817} is T-dual to the $\sigma$-model \eqref{818}. } 
 
 \medskip
 
\noindent The proof of {\bf Assertion 3} is in every aspect similar to the one of Assertion 2 in Section 5, we shall be  therefore brief.  We first remark that it holds
 \be E^\qa\ \!^R\nabla^l=E^{\aq}\ \!^L\nabla^l=\ \!^\qa{\bm\nabla}^l,\label{879}\ee
 where the operators $\ \!^L\nabla^l$ and $\ \!^R\nabla^l$ where defined in \eqref{824} and we define 
 the $\K$-valued differential operator $\ \!^\qa{\bm\nabla}^l$ acting on the functions on $K$ as
 \be \left(\ \!^\qa{\bm\nabla}^l f, x\right)_\qa :=  \nabla^l_{x} f, \qquad  x\in \K.\label{882}\ee
 To see e.g. that the first of the relations \eqref{879} indeed holds, we rewrite the left-hand-side of the second of Eqs. \eqref{824}
 as 
 \be \left(\ \!^\qa{\bm\nabla}^l f, x\right)_\qa:= \nabla^l_{x} f = \left(\ \!^{R}\nabla^l f, x\right)_{\D_R}=\left({E^\na}^{-1}E^\na\ \!^{R}\nabla^l f, x\right)_{\D_R}=
 \left(E^\qa\ \!^{R}\nabla^l f, x\right)_\qa \ee
 and  we finish the argument by invoking the non-degeneracy of the bilinear forms $(.,.)_{\D_R}$ and $(.,.)_\qa$.
 
 We shall need also the following relation
 \be \Pi^{\aq}(k)= Ad_k\Pi^\qa(k^{-1})({E^\qa})^{-1}Ad_{k^{-1}}E^{\aq},\label{890}\ee
 which can be obtained from the identity \eq{876} much in the same way as the relation \eq{cru} was obtained from the identity \eq{331} in Section 5.2.
 
 We can finally 
 rewrite the $\sigma$-model actions \eqref{817} and \eqref{818} as follows
$$ S_\qa(k)= \jp\int d\tau \oint \biggl({E^\qa}^{-1}\Bigl(1+\Pi^\qa(k){E^\qa}^{-1}\Bigr)^{-1}\partial_+k k^{-1}, \partial_- k k^{-1}\biggr)_{\D_R} =$$\be =\jp\int d\tau \oint \biggl(\Bigl(1+\Pi^\qa(k){E^\qa}^{-1}\Bigr)^{-1}\partial_+k k^{-1}, \partial_- k k^{-1}\biggr)_\qa\label{894}\ee
 $$ S_{\aq}(k)= \jp\int d\tau \oint \biggl({E^{\aq}}^{-1}\Bigl(1+\Pi^{\aq}(k){E^{\aq}}^{-1}\Bigr)^{-1}\partial_+k k^{-1}, \partial_- k k^{-1}\biggr)_{\D_L} =$$\be =\jp\int d\tau \oint \biggl(\Bigl(1+\Pi^{\aq}(k){E^{\aq}}^{-1}\Bigr)^{-1}\partial_+k k^{-1}, \partial_- k k^{-1}\biggr)_{\aq}.\label{895}\ee
 Using the crucial identity \eqref{890} as well as the hypothesis that the ad$_\K$-invariant bilinear forms   $(.,.)_\qa$ and $(.,.)_{\aq}$ coincide, we find finally
 \be S_\qa(k)=S_{\aq}(k^{-1}).\label{898}\ee
We conclude that if the sufficient condition of {\bf Assertion 3} holds, then the  T-duality between the $\sigma$-models 
 \eqref{817} and \eqref{818} is   realized  simply
 by the  field redefinition $k\to k^{-1}$  and the corresponding T-duality symplectomorphism is therefore just the point canonical transformation. On the other hand, the affine quasi-Poisson  T-duality between the models \eq{820} and \eq{822} is nontrivial because it is the composition of two horizontal and one (left) vertical symplectomorphism  in the scheme \eq{Sch2} and only the left vertical one is point-like.
 
 \subsection{Examples of the affine quasi-Poisson T-duality} 

 If the affine quasi-Poisson group $(K,\pi^\qa)$ and the operators $E^\qa,E^{\aq}$ satisfy the 
  sufficient condition of {\bf Assertion 3} then the Lie algebra $\K$ is quadratic, which means that  there is the symmetric ad$_\K$-invariant
  non-degenerate bilinear form $(.,.)_\K$ on  $\K$. Let us therefore restrict our effort to construct affine quasi-Poisson structures  to the case of the groups $K$ which have quadratic Lie algebras $\K$.
   We first define two Lie quasi-bialgebra structures on $\K$ and then construct the affine quasi-Poisson structure $\pi^\qa$ compatible with them in the sense of the conditions  \eq{pmc}, \eq{pmd} and \eq{qp6}.
   
   Let $R^r:\K\to\K$ and $R^l:\K\to\K$ be two Yang-Baxter operators, which means that they are both skew-symmetric with respect to the bilinear form $(.,.)_\K$ and they verify the Yang-Baxter identities with different values of the parameter $\e$:
    \be [R^rx,R^ry]=R^r([R^rx,y]+[x,R^ry])-\e_r[x,y], \qquad \forall x,y\in \K; \label{913}\ee 
  \be [R^lx,R^ly]=R^l([R^lx,y]+[x,R^ly])-\e_l[x,y], \qquad \forall x,y\in \K. \label{914}\ee 
  Consider also an element $\chi\in\Lambda^3\K$ defined by the relation
  \be (\chi,x\otimes y\otimes z)_{\K\otimes\K\otimes\K}\equiv (\chi',x)_\K(\chi'',y)_\K(\chi''',z)_\K:=(\e_r-\e_l)([x,y],z)_\K,\quad x,y,x\in\K,\label{916}\ee
  where we have used the Sweedler notation $\chi=\chi'\otimes\chi''\otimes\chi'''$.
  The definition of the affine quasi-Poisson structure requires   the element $\chi$  to be $\K$-invariant, which is indeed the case because
 the ad$_\K$-invariance of $\K$ implies that it holds for every $u,x,y,z\in\K$ 
  \be ([ad_ux,y],z)_\K+([x,ad_uy],z)_\K+([x,y],ad_uz)_\K= 0.\ee
  Let us now identify the spaces $\K^*$ and $\K$ via the non-degenerate form
  $(.,.)_\K$, that is, we have the identification map $J:\K^*\to\K$ defined as 
  \be (J(\alpha),x)_\K=\la \alpha,x\ra.\ee
We define  the left Lie quasi-bialgebra structure on $\K$ as the quadruple
$\K_L:=(\K,[.,.],[.,.]_L^*,\chi)$ where the $3$-form $\chi\in\Lambda^3\K$ is given by 
Eq.\eq{916} and the dual commutator $[.,.]^*_L$ on $\K^*$ is given by
\be [\alpha,\beta]^*_L:=-J^{-1}\left([R^lJ(\alpha),\beta]+[J(\alpha),R^lJ(\beta)]\right).\label{928}\ee
Similarly, the  right Lie quasi-bialgebra structure on $\K$ is the quadruple
$\K_R:=(\K,[.,.],[.,.]_R^*,-\chi)$ where   the dual commutator $[.,.]^*_R$ on $\K^*$ is given by
\be [\alpha,\beta]^*_R:=-J^{-1}\left([R^rJ(\alpha),\beta]+[J(\alpha),R^rJ(\beta)]\right).\label{932}\ee
Before defining the affine quasi-Poisson structure $\pi^\qa$ compatible with the actions of the two Lie quasi-bialgebras just introduced, we make a little digression and try to understand the structure of the respective  quasi-Drinfeld doubles $D^q_L$ and $D^q_R$ of the Lie quasi-bialgebras $\K_L$ and $\K_R$. 
We start with $\D^q_L$ and remark that the direct application of the general formula \eqref{qdc} gives the following result
$$ [x\oplus\alpha, y\oplus \beta]_{\D^q_L}=\left([x,y]-[x,R^lJ(\beta)]+R^l[x,J(\beta)]+[y,R^lJ(\alpha)]-R^l[y,J(\alpha)] +(\e_r-\e_l)[J(\alpha),J(\beta)]\right)\oplus$$\be \oplus\left(-J^{-1}[J(\alpha),J(\beta)]^*-J^{-1}[J(\beta),x]+J^{-1}[J(\alpha),y]\right).\label{934}\ee
This formula looks quite cumbersome but there is a simple isometric isomorphism $\ups_L$ from $\D^q_L$ into the Lie algebra $\D_{\e_r}$ with the commutator defined by Eq.\eq{str}
\be [(x_1,x_2)_r,(y_1,y_2)_r]_{\e_r}=([x_1,y_1]+\e_r[x_2,y_2],[x_1,y_2]+[x_2,y_1])_r.\label{938}\ee
The map $\ups_L$ is given by
\be \ups_L(x\oplus \al)=\left(x-R^lJ(\alpha),J(\alpha)\right)_r\ee
and the Yang-Baxter property of the operator $R^l$ is the only thing needed for proving that the map $\ups_L$ is indeed the isomorphism of the  Lie algebras $\D^q_L$ and $\D_{\e_r}$. Similarly, 
the map $\ups_R$  given by
\be \ups_R(x\oplus \al)=\left(x-R^rJ(\alpha),J(\alpha)\right)_l\ee
realizes the isometric  isomorphism of the  Lie algebras $\D^q_R$ and $\D_{\e_l}$.

We notice that the structures of the quasi-doubles $\D^q_L$ and $\D^q_R$ of the Lie quasi-bialgebras $\K_L$ and $\K_R$ do not depend on the particular form of the operators $R^l$ and $R^r$ but only on the parameters $\e_r$ and $\e_l$.
On the other hand, the ad-invariant non-degenerate symmetric bilinear forms $(.,.)_{\D_{\e_r}}$   and $(.,.)_{\D_{\e_l}}$  do not depend even on the real parameters $\e_r,\e_l$ and they are given by  the
formulae
\be ((x_1,x_2)_r,(y_1,y_2)_r)_{\D_{\e_r}}:= (x_2,y_1)_\K+(x_1,y_2)_\K.\label{918}\ee
\be ((x_1,x_2)_l,(y_1,y_2)_l)_{\D_{\e_l}}:= (x_2,y_1)_\K+(x_1,y_2)_\K.\label{919}\ee
We also note that the  Lie algebra $\K$ is embedded into $\D_{\e_r}$ and into $\D_{\e_l}$ in the same way, i.e.  $(\K,0)\subset \D_{\e_r(\e_l)}$.

 The affine quasi-Poisson bracket $\{.,.\}^\qa$ compatible with the left and right actions of the Lie quasi-bialgebras $\K_L$ and $\K_R$ is then given by 
  \be \{f_1,f_2\}^\qa_K:=({\bm\nabla}^r f_1,R^r{\bm\nabla}^r f_2)_\K+({\bm\nabla}^l f_1,R^l{\bm\nabla}^l f_2)_\K,\label{955}\ee
  which  
 the $\K$-valued differential operators ${\bm\nabla}^r, {\bm\nabla}^l$ acting on the functions on $K$ were defined in Eq.\eq{bft'}.
  The formula \eq{955} looks identical as in the affine Poisson case \eq{aff} but now the  operators $R^l,R^r$  verify the Yang-Baxter equations with different values of the parameters $\e_l\neq\e_r$. In consequence, the bracket \eq{955} is not Poisson but just quasi-Poisson. Let us indeed verify that the bracket
  \eq{955} is affine quasi-Poisson  according to the defining relations 
   \eq{pmc}, \eq{pmd} and \eq{qp6}.
   
 In order to verify the conditions \eq{pmc} and \eq{pmd}, we choose the orthonormal basis 
 $t_i$ on $\K$ and write the bivector $\pi^\qa$ corresponding to the bracket \eq{955} as
 \be \pi^\qa=\jp(t_i,R^rt_j)\nabla_{t_i}^r\wedge\nabla^r_{t_j}+\jp(t_i,R^lt_j)\nabla_{t_i}^l\wedge\nabla^l_{t_j}.\ee
 Since the left-invariant vector fields on $K$ commute with the right-invariant ones,
 we find, respectively, the desired results
 \be \L_{\nabla^l_{t_k}}\pi^\qa= -\jp(t_i,R^lt_j)_\K\left(\nabla_{[t_k,t_i]}^l\wedge\nabla^l_{t_j}+\nabla_{t_i}^l\wedge\nabla^l_{[t_k,t_j]}\right)=
 \jp\left([t_i,R^lt_j]+[R^lt_i,t_j],t_k\right)_\K          \nabla_{t_i}^l\wedge \nabla_{t_j}^l.\label{966}\ee
 \be \L_{\nabla^r_{t_k}}\pi^\qa=\jp (t_i,R^rt_j)\left(\nabla_{[t_k,t_i]}^r\wedge\nabla^r_{t_j}+\nabla_{t_i}^r\wedge\nabla^r_{[t_k,t_j]}\right)=-\jp\left([t_i,R^rt_j]+[R^rt_i,t_j],t_k\right)_\K          \nabla_{t_i}^r\wedge \nabla_{t_j}^r.\label{967} \ee 
 because the dual structure constants  $\ ^{L}\tilde c_i^{\ jk}$ and
 $\ ^{R}\tilde c_i^{\ jk}$ featuring in \eq{pmc} and \eq{pmd} are given by the dual commutators
 \eq{928} and \eq{932} of the elements of the basis $t_i$.
 
 It remains to calculate the Schouten bracket 
 $$ \jp[\pi^\qa,\pi^\qa]_K=$$\be=\frac{1}{8}(t_i,R^rt_j)_\K(t_k,R^rt_m)_\K \left[\nabla_{t_i}^r\wedge \nabla_{t_j}^r, \nabla_{t_k}^r\wedge \nabla_{t_m}^r\right]_K +\frac{1}{8}(t_i,R^lt_j)_\K(t_k,R^lt_m)_\K \left[\nabla_{t_i}^l\wedge \nabla_{t_j}^l, \nabla_{t_k}^l\wedge \nabla_{t_m}^l\right]_K.\label{973} \ee
 We use  the formula \eq{247} and the Yang-Baxter identities \eq{913} and \eq{914}  to obtain the desired result
$$ \jp[\pi^\qa,\pi^\qa]_K=-\frac{1}{6}\left([R^lt_i,R^lt_j]-R^l\left([R^lt_i,t_j]+[t_i,R^lt_j]\right),t_k\right)_\K \nabla_{t_i}^l\wedge \nabla_{t_j}^l\wedge\nabla_{t_k}^l +$$\be +\frac{1}{6} \left([R^rt_i,R^rt_j]-R^r\left([R^rt_i,t_j]+[t_i,R^rt_j]\right),t_k\right)_\K \nabla_{t_i}^r\wedge \nabla_{t_j}^r\wedge\nabla_{t_k}^r=\frac{1}{6}(\e_r-\e_l)\left([t_i,t_j],t_k\right)_\K \nabla_{t_i}^r\wedge \nabla_{t_j}^r\wedge\nabla_{t_k}^r.\label{975}
  \ee
The  mirror affine quasi-Poisson bracket $\{.,.\}^{\aq}$ and the associated left and write Poisson-Lie brackets  $\{.,.\}^L$,  $\{.,.\}^R$ can be easily extracted from Eqs. \eqref{717}  and \eqref{760} and they are given by the formulae
\be \{f_1,f_2\}^{\aq}_K:=({\bm\nabla}^r f_1,R^l{\bm\nabla}^r f_2)_\K+({\bm\nabla}^l f_1,R^r{\bm\nabla}^l f_2)_\K,\label{978}\ee
\be \{f_1,f_2\}^{L}_K:=({\bm\nabla}^r f_1,R^l{\bm\nabla}^r f_2)_\K-({\bm\nabla}^l f_1,R^l{\bm\nabla}^l f_2)_\K,\label{979}\ee
\be \{f_1,f_2\}^{R}_K:=({\bm\nabla}^r f_1,R^r{\bm\nabla}^r f_2)_\K-({\bm\nabla}^l f_1,R^r{\bm\nabla}^l f_2)_\K.\label{980}\ee
The fact that the brackets $\{.,.\}^L$ and  $\{.,.\}^R$ are Poisson-Lie can be also verified by a direct calculation which amounts to replacing appropriately $R^r$,$R^l$ in  Eqs. \eq{966}, \eq{967}, \eq{973} and \eq{975}. For example, for $\pi^L$ we replace $R^r$ by $R^l$ and $R^l$ by $-R^l$ and the result is
\be \L_{\nabla^l_{t_k}}\pi^L =-\jp
 \left([t_i,R^lt_j]+[R^lt_i,t_j],t_k\right)_\K          \nabla_{t_i}^l\wedge \nabla_{t_j}^l;\label{981}\ee
 \be \L_{\nabla^r_{t_k}}\pi^L =-\jp\left([t_i,R^lt_j]+[R^lt_i,t_j],t_k\right)_\K          \nabla_{t_i}^r\wedge \nabla_{t_j}^r;\label{982} \ee 
\be [\pi^L,\pi^L]_K =(\e_l-\e_l)\left([t_i,t_j],t_k\right)_\K \nabla_{t_i}^r\wedge \nabla_{t_j}^r\wedge\nabla_{t_k}^r=0.\label{983}
  \ee
  Therefore $\pi^L$ and, similarly, $\pi^R$ are indeed the Poisson-Lie structures.
  
  What are the Drinfeld doubles $D_L$ and $D_R$  for the Poisson-Lie groups
  $(K,\pi^L)$ and $(K,\pi^R)$? Well, these are  the doubles $D_{\e_l}$ and $D_{\e_r}$, respectively, where the corresponding maximally isotropic Lie subalgebras $\tilde \K_L$ and $\tilde \K_R$ are given as the graphs of the operators $R^l$ and $R^r$:
\be \tilde\K_L =\{(-R^lx,x)_l, x\in\K\},\quad  \tilde\K_R =\{(-R^rx,x)_r, x\in\K\}.\label{990}\ee
 It can be checked that the explicit formulae \eq{plr} and \eq{314} applied to the subspaces
 \eq{990} reproduce the formulae \eq{979} and \eq{980} as they should. 
 
 Can the affine quasi-Poisson structure \eq{955} serve as the basis for a viable example of the affine quasi-Poisson T-duality? The answer to this question is affirmative because it is   easy to verify  that in this case the sufficient condition of {\bf Assertion 3} holds. Indeed, the following choice of
  the operators 
$E^\qa:\tilde\K_R\to\K$ and $E^{\aq}:\tilde\K_L\to\K$ does the job
\be E^\qa(-R^rx,x)_r:=a(x,0)_r, \qquad  E^{\aq}(-R^lx,x)_l:=a(x,0)_l, \qquad a<0.\label{994}\ee
Using the definitions \eqref{918} and \eq{919},  it is then easy to verify for every $x,y\in \K$ that there hold  the equalities of the following  inner products
\be (x,y)_\qa\equiv ((x,0)_r,(E^\qa)^{-1}(y,0)_r)_{D_{\e_r}} =\frac{1}{a}((x,0)_r,(-R^ry,y)_r)_{\D_{\e_r}}=\frac{1}{a}(x,y)_\K,\ee
\be (x,y)_{\aq}\equiv ((x,0)_l,(E^{\aq})^{-1}(y,0)_l)_{\D_{\e_l}}=\frac{1}{a}((x,0)_l,(-R^ly,y)_l)_{\D_{\e_l}}=\frac{1}{a}(x,y)_\K,  \ee
hence the sufficient condition for the affine quasi-Poisson T-duality is indeed satisfied.

It is  instructive to cast the $\sigma$-model actions \eqref{817} and  \eqref{818}
in terms of the Yang-Baxter operators $R^l,R^r$. Similarly as in Section 5.3, we    find
\be \Pi^\qa(k)({E^\qa})^{-1}=\frac{1}{a}R^l+\frac{1}{a}R^r_{k^{-1}}, \quad \Pi^{\aq}(k)({E^{\aq}})^{-1}=\frac{1}{a}R^r+\frac{1}{a}R^l_{k^{-1}},\label{ger'}\ee
where 
\be R^l_{k^{-1}}:=Ad_k R^l Ad_{k^{-1}}, \quad  R^r_{k^{-1}}:=Ad_{k} R^r Ad_{k^{-1}}.\ee 
Finally,  we infer
\be S_\qa(k)=\jp\int d\tau \oint \left(\left(a+  R^r+R^l_{k^{-1}}\right)^{-1}\partial_+k k^{-1}, \partial_- k k^{-1}\right)_\K;\label{008}\ee
\be S_{\aq}(k)=\jp\int d\tau \oint \left(\left(a+ R^l+R^r_{k^{-1}}\right)^{-1}
\partial_+k k^{-1}, \partial_- k k^{-1}\right)_\K.\label{009}\ee
The affine quasi-Poisson formulae \eq{008} and \eq{009} seem to coincide with the affine Poisson ones \eq{kad} and \eq{kbd}, but the difference resides in the fact that in the affine quasi-Poisson case the operators $R^l$ and $R^r$   verify the Yang-Baxter conditions \eq{913} and \eq{914} with $\e_l\neq\e_r$.

Of course, there is no universal way to rewrite the actions of the dual $\sigma$-models $\tilde S_{\qa}(\tilde k_R)$ and $\tilde S_{\aq}(\tilde k_L)$
in terms of the Yang-Baxter operators $R^r,R^l$ since the very structure of the groups $\tilde K_L,\tilde K_R$ depends implicitely  on $R^r,R^l$. The formulae \eq{820} and \eq{822} represent the maximum that can be achieved in this direction, on the other hand, the operators $R^r$ and $R^l$ are  useful if we wish to describe explicitely the subspaces $\E_L\subset \D_{\e_l}$ and $\E_R\subset\D_{\e_r}$ underlying, respectively, the Poisson-Lie T-dualities relating the model \eqref{008} with the model \eqref{820} and the model \eqref{009} with the model \eqref{822}. We find with the help of the formulae  \eqref{843} and \eqref{ger'} that
\be M^R(E^\qa)^{-1}= M^L(E^{\aq})^{-1}=\frac{1}{a}R^r+\frac{1}{a}R^l,\label{014}\ee
hence
 \be \E_R= \{(ay+R^ly,y)_r,y\in\K\},\quad  \E_L= \{(ay+R^ry,y)_l,x\in\K\}.\label{017}\ee

We conclude this section by listing several distinguished choices of the Yang-Baxter operators $R^l,R^r$ leading to the affine quasi-Poisson T-duality. We first consider the case $R^l=0$ and $R^r$ arbitrary  in which case the action \eq{008} is that of the Yang-Baxter $\sigma$-model introduced by the present author in \cite{K02,K09}. This case is distinguished by the fact that  the Drinfeld doubles
$\D_L$ and $\D_R$ are not isomorphic.

If one of the operators $R^l, R^r$ is a multiple of the other then the corresponding action \eq{008} describes another particular $\sigma$-model introduced by the present author in \cite{K09,K14} in connection with the question of integrability of nonlinear $\sigma$-models. Finally, we may consider 
an affine quasi-Poisson generalisation  of the affine Poisson T-duality based on the Drinfeld twist operators (cf. Section 4.4). In this
case we have \be R^l=\alpha_l R^c+R_t^l,\quad  R^r=\alpha_r R^c+R_t^r, \quad \alpha_l,\alpha_r\in\br,\ee
where $R_t^l$, $R_t^r$ are the Drinfeld twist operators introduced at the beginning of Section 4.4 and  $R^c:\K\to\K$ is the so-called canonical Yang-Baxter operator defined as (cf. Ref. \cite{K09}):
\be R^cT^\mu=0,\quad R^cB^\al=C^\al,\quad R^cC^\al=-B^\al.\label{022}\ee
Here $T^\mu$ is a basis of the Cartan subalgebra $\gH$ of the real simple compact Lie algebra $\K$ and
the basis of $\gH^\perp\subset \K$
is chosen as $ B^{\al},C^{\al}$,
 $\al>0$ where 
\be   B^{\al}=\frac{i}{\sqrt{2}}(E^{\al}+E^{-\al}),\quad 
C^{\al}=\frac{1}{\sqrt{2}}(E^{\al}-E^{-\al})\ee
and $E^{\pm\al}$ are the step generators of $\K^\bc$.

 It is not difficult to check that the skew-symmetric operator $R^c$ verifies  the  Yang-Baxter identity 
  \be [R^cx,R^cy]=R^c([R^cx,y]+[x,R^cy])+[x,y], \quad x,y\in \K \label{II}\ee
 and, using this fact, to verify that  $R^l=\alpha_l R^c+R_t^l$ and $R^r=\alpha_r R^c+R_t^r$ are also the Yang-Baxter operators with $\e_l=-\alpha_l^2$ and $\e_r=-\alpha_r^2$, respectively. 

\section{Dressing cosets}
There exists a generalization of the Poisson-Lie T-duality  introduced as the
Poisson-Lie counterpart of the standard non-Abelian T-duality for the cases where the non-Abelian isometry group does not act freely on the target space \cite{KS96b}. This so-called  "dressing cosets" construction gives rise seemingly to the same dual pair of the $\sigma$-models 
\eqref{mmm} and \eqref{mmm'} as the standard Poisson-Lie T-duality but the linear operators $E,\tilde E$ featuring in the Lagrangians are such that the both  models
\eqref{mmm} and \eqref{mmm'}  develop a gauge symmetry reducing the common dimension of their targets. In particular, if the target groups $K$ and $\tilde K$ are $d$-dimensional and the dimension of the gauge group $F$ is $p$, then after the gauge fixing the  targets of the $\sigma$-models \eqref{mmm} and \eqref{mmm'} become effectively $(d-p)$-dimensional.

As an example of the dressing cosets construction, consider the dual pair of the $\sigma$-models \eqref{mmm} and \eqref{mmm'} for the case when $K$ is a simple compact group and  $\tilde K$ is the Lie algebra $\K$ with the (Abelian) group structure given by the vector space addition. The Drinfeld double is the group $D_0$ with the multiplication law given by Eq. \eqref{gl}, the Poisson-Lie bivector $\Pi(k)$ on $K$ then trivally vanishes and the Poisson-Lie bivector on $\tilde K$  is given by the adjoint action of the Lie algebra. The actions of the $\sigma$-models \eqref{mmm} and \eqref{mmm'} in this particular case thus become
\be S=\jp\int d\tau\oint d\sigma     \biggl( \tilde E\partial_+k k^{-1}, \partial_- k k^{-1}\biggr)_\D, \qquad k(\tau,\sigma)\in K;\label{1131}\ee 
\be \tilde S=\jp\int d\tau\oint d\sigma    \biggl( \Bigl(\tilde E- {\rm ad}_\kappa \Bigr)^{-1}\partial_+\kappa  , \partial_- \kappa \biggr)_\D, \qquad \kappa(\tau,\sigma)\in \K.\label{1132}\ee 
For the linear operator $\tilde E$ we pick the orthogonal projector on the subspace $\gH^\perp\subset \K$ perpendicular to the Cartan subalgebra $\gH$. Then  it is not  difficult to see that both $\sigma$-models \eqref{1131} and \eqref{1132} develop
the gauge symmetry with respect to the action of the Cartan torus $\mathbb T\subset K$ (the Lie algebra of the Cartan torus is the Cartan subalgebra $\gH$). In particular, an element $f(\tau,\sigma)$ of the gauge group acts as
\be k\to fk, \qquad \kappa\to {\rm Ad}_f\kappa.\ee
For example, for the group $K=SU(2)$ the models  \eqref{mmm} and \eqref{mmm'}  have both two-dimensional targets and correspond
to the dual pairs obtained by different methods already in the early days of the non-Abelian T-duality \cite{OQ92,GR93,AABL93,Hew96}. Other examples of the dressing cosets have been studied in \cite{S98,S99,KP99, CM06, HT, SST15, BTW, HS,SV}. 

The dressing cosets method  may  not look like a substantial generalisation of the Poisson-Lie T-duality, because it boils down just to the study of some special linear operators $E$ for which the dual $\sigma$-models \eqref{mmm} and \eqref{mmm'}  develop the gauge symmetry. It is, however, the very purpose of the present section to show that the dressing cosets generalization of the Poisson-Lie T-duality
is not   as mild one as it may seem, since it covers all examples of the affine (quasi-)Poisson T-duality which we have constructed in the previous sections! 

A remark is in order at this point of the exposition. When we say that the affine (quasi-)Poisson generalization of the Poisson T-duality is already included in the dressing cosets construction \cite{KS96b}, we do not mean that we have exposed 
in the previous sections  the story which had been already known.
Actually, the existence of the pairwise T-duality of the four $\sigma$-models \eqref{mm1}, \eqref{mm2}, \eqref{mm3}, \eqref{mm4} associated to the affine (quasi-)Poisson geometry constitutes the genuinely new result of the present article. Of course, we could  have also  presented this result as a new nontrivial application of the old dressing cosets method but this would not respect the logic of the things.  Indeed, in reality, we have discovered the affine (quasi-)Poisson T-duality  by reasoning making no reference to the dressing cosets and only thanks to the insights obtained in this way we were able to find out {\it \`a posteriori}  that there is beyond all that an appropriate variant of the dressing cosets construction.  It seems difficult to imagine  to move in the opposite direction. Indeed, it looks counterintuitive to suspect that by  taking particular singular 
values of the operators $E,\tilde E$ in the actions 
\eqref{mmm} and \eqref{mmm'} the quadruple of the four pairwise T-dual  $\sigma$-models \eqref{mm1}, \eqref{mm2}, \eqref{mm3}, \eqref{mm4} could emerge, and it neither seems evident, how the dressing cosets construction could explain that two $\E$-models living on different Drinfeld doubles  can be dynamically equivalent. Nevertheless, it is precisely what eventually happens... 

To move on forward, we have first to review the first order formalism of the dressing cosets. As in the case of the standard Poisson-Lie T-duality, the construction is underlied by a particular Drinfeld double which we now denote as $\mathbb D$. Three subgroups of the double $\dd$  enter into the game: two of them are the half-dimensional isotropic subgroups $K$ and $\tilde K$ as before but there is also a subgroup $F$, which is isotropic too (i.e. the restriction of the bilinear form $(.,.)_{{\rm Lie}(\dd)}$ on the Lie algebra 
$\F$ of the group $F$ vanishes). The group $F$ may have whatever dimension smaller than the half of the dimension of the double and plays the role of the gauge group.

The
phase space $L\dd_F$ of the dressing coset $\E$-model is now the space of the elements $l(\sigma)$ of the loop group $L\dd$, for which it holds
\be (\d_\sigma ll^{-1},\F)_{{\rm Lie}(\dd)}=0.\ee
The (pre)symplectic form of the dressing coset $\E$-model is just the restriction to  $L\dd_F$ of the symplectic form 
\be \omega_{L\mathbb D}=-\jp \oint d\sigma  (l^{-1}dl,\partial_\sigma(l^{-1}dl))_{{\rm Lie}(\dd)}\label{1158}\ee
and the Hamiltonian of the dressing coset $\E$-model looks the same as in the case of the standard Poisson-Lie T-duality  
\be Q_\E=\jp\oint d\sigma  (\partial_\sigma ll^{-1},\E \partial_\sigma ll^{-1})_{{\rm Lie}(\dd)}.\label{1160}\ee
However, now  the linear operator $\E$ has different properties. Denoting by $\F^\perp$ all elements in the Lie algebra  Lie$(\dd)$ orthogonal to $\F$, we require that the operator
$\E:\F^\perp\to\F^\perp$ is self-adjoint, its kernel contains 
$\F$, the bilinear form $(.,\E.)_{{\rm Lie}(\dd)}$ on $\F^\perp$ is positive semi-definitive, the image of the operator $\E^2-{\rm Id}$ is contained in $\F$ and, finally, 
$\E$ must commute with the adjoint action of the Lie algebra $\F$ on the vector space $\F^\perp$.

By decomposing the elements $l(\sigma)$ of the phase space $L\dd_F$ in two ways as $l=k\tilde h$ and $l=\tilde kh$ and by eliminating the
fields $\tilde h$ and $h$, we obtain, respectively, the $\sigma$-models \eqref{mmm} and \eqref{mmm'}, where the linear operators $E$ and $\tilde E$ are obtained by decomposing the image of the operator  $\E+{\rm Id}$ as
\be {\rm Im}(\E+{\rm Id})=\{\tilde x +E\tilde x,\tilde x\in\tilde\K\}=\{  x +\tilde E x, x\in\K\}.\ee
The reader can find the details of this procedure  in the original paper \cite{KS96b}.

In order to produce  the affine (quasi-)Poisson T-duality scheme \eqref{Sch2} out from the formalism of the dressing cosets, we choose the Drinfeld double ${\mathbb D}$ of the following form
\be \mathbb D=D_{\e_l}\times D_{\e_r}\ee
where $D_{\e_l}$ and $D_{\e_r}$ are two Drinfeld doubles of the group $K$ defined by Eq.\eqref{str}.
The invariant split bilinear form $(.,.)_{{\rm Lie}(\dd)}$ is given by
\be ((u_l,u_r),(v_l,v_r))_{{\rm Lie}(\dd)}:=(u_l,v_l)_{\D_{\e_l}}+(u_r,v_r)_{\D_{\e_r}}, \qquad  u_l,v_l\in {\D_{\e_l}}, \quad u_r,v_r\in {\D_{\e_r}}\ee
and the correct choice of the gauge Lie subalgebra $\F\subset$ Lie $(\mathbb D)$ is 
\be \F=\{(x,x),\quad  x\in\K\}.\label{1177}\ee
Detailing the elements $u_l\in{\D_{\e_l}}$ and $u_r\in{\D_{\e_r}}$
as
\be u_l=(x_l,y_l), \quad  u_r=(x_r,y_r), \qquad x_l,y_l,x_r,y_r\in \K,\ee we  first rewrite \eqref{1177} as 
\be \F=\Bigl\{\bigl((x,0),(x,0)\bigr), x\in \K \Bigr\}\ee
and then we find the subspace $\F^\perp$:
\be \F^\perp=\Bigl\{\bigl((x_l,y),(x_r,-y)\bigr), x_l,x_r,y\in \K \Bigr\}.\ee
It remains to choose the operator $\E:\F^\perp\to\F^\perp$. In turns out, that the following one  has all needed properties and does the job 
\be \E\bigl((x_l,y),(x_r,-y)\bigr):=\bigl((y,\jp(x_l-x_r)),(-y,\jp(x_r-x_l))\bigr).\ee
How do we obtain the four $\sigma$-models \eqref{mm1}, \eqref{mm2}, \eqref{mm3} and \eqref{mm4} from the first order formulae 
\eqref{1158} and \eqref{1160}, knowing  that the standard dressing cosets construction gives rise just to the two $\sigma$-models \eqref{mmm} and \eqref{mmm'}? Well, the key trick  is to use here the construction known as the Poisson-Lie T-plurality \cite{KS97,KS97b,vU02} which amounts to the statement that there are as many dynamically
equivalent $\sigma$-models with different geometries extracted from the first order formalism \eqref{1158} and \eqref{1160}, as is the number of maximally isotropic subgroups of the Drinfeld double
$\dd$ not related by internal automorphisms. If there
are just two maximally isotropic subgroups $K$ and $\tilde K$, then there are just two mutually dual $\sigma$-models \eqref{mmm} and \eqref{mmm'}. However, our double $\dd=D_{\e_l}\times D_{\e_r}$ has more than two maximally isotropic subgroups; we shall actually need three of them: $\tilde K_L\times \tilde K_R$, $\tilde K_L\times K$ and $K\times \tilde K_R$, where
$\tilde K_L\subset D_{\e_l}$ and $\tilde K_R\subset D_{\e_r}$ are the maximally isotropic subgroups of the respective doubles 
$D_{\e_l}$ and $D_{\e_r}$ (their Lie algebras are described by Eq. \eqref{990}). 

Following the general dressing cosets construction \cite{KS96b},
the triple of the pairwise T-dual $\sigma$-models have for  their respective targets the double cosets $F\backslash \dd/(\tilde K_L\times \tilde K_R)$,
$F\backslash \dd/(\tilde K_L\times K)$ and $F\backslash \dd/(K\times \tilde K_R)$ or, equivalently,
$F\backslash (K\times K)$, $F\backslash (K\times \tilde K_R)$ and $F\backslash (\tilde K_L\times K)$. Furthermore, the gauge group
$F$ is isomorphic to $K$, therefore the targets $F\backslash (K\times \tilde K_R)$ and $F\backslash (\tilde K_L\times K)$ become
simply $\tilde K_R$ and $\tilde K_L$ and a straightforward computation shows that the $\sigma$-model living on those targets are nothing but the $\sigma$-models \eqref{mm3} and \eqref{mm4}.
Concerning the remaining target $F\backslash (K\times K)$, we can fix the gauge in two different ways: either we obtain the target $\{e_K\}\times K$ or $K\times \{e_K\}$, where $e_K$ stands for the unit element of the group $K$. Those two gauge choices turn out to result at the $\sigma$-models \eqref{mm1} and \eqref{mm2} living on the target $K$. All in all, the affine (quasi-)Poisson T-duality of the four $\sigma$-models \eqref{mm1}, \eqref{mm2}, \eqref{mm3} and \eqref{mm4} fits into the dressing cosets construction.

\section{Renormalisation group flow and the Drinfeld twist operators}
 \setcounter{equation}{0}

It was discovered  in \cite{VKS,SfS} that the ultraviolet corrections to the actions of the Poisson-Lie $\sigma$-models \eqref{mmm} and \eqref{mmm'} can be absorbed by appropriate  redefinitions of the linear operators $E$ and $\tilde E$. Moreover, those redefinitions respect the T-duality between the $\sigma$-models. Said in other words, the renormalisation group flow  of the operators $E$ and $\tilde E$ calculated separately from the models \eqref{mmm} and \eqref{mmm'} respects
the duality condition requiring that $E$ is inverse to $\tilde E$. Actually, the RG flow of the pair of the mutually dual $\sigma$-models \eqref{mmm} and \eqref{mmm'} can be calculated also from the duality invariant data $(LD,\omega_{LD}, Q_\E)$  introduced in Section 2. The quantity which flows in the duality invariant description is the subspace $\E$, and this flow was described by the following elegant formula derived in \cite{SST}:
\be \frac{d\E_{AB}}{ds}=k\left(\E_{AC}\E_{BF}-\eta_{AC}\eta_{BF}\right)\left(\E^{KD}\E^{HE}-\eta^{KD}\eta^{HE}\right)f_{KH}^{\ \ \ \ \!C}f_{DE}^{\ \ \ \ \!F}. \label{Sf}\ee
Here $s$ is the flow parameter, $k$ a constant, the capital Latin indices refer to the choice of a basis $T_A$ in the Lie algebra $\D$:
\be \E_{AB}:=(T_A,\E T_B)_\D, \quad \eta_{AB}:=(T_A,T_B)_\D,\quad [T_A,T_B]=f_{AB}^{\phantom{AB}C}T_C\ee
and they are respectively lowered  and raised with the help of the tensor $\eta_{AB}$ and its inverse. Recall also,  that   $\E$ is the self-adjoint linear operator 
$\E:\D\to\D$ which has the subspace $\E$ as the eigenspace for the eigenvalue $+1$ and the orthogonal complement subspace $\E^\perp$  as the eigenspace for the eigenvalue $-1$.  

Up to an irrelevant normalization constant, the flow formula \eqref{Sf} can be cast in the basis-independent way as follows:
\be \frac{d\E}{ds}=\P_+[[\P_+,\P_-]]\P_-+\P_-[[\P_+,\P_-]]\P_+,\label{flo}\ee
where the operators $\P_\pm$ are defined as
\be \P_\pm=\jp(1\pm \E),\ee
and the double bracket $[[.,.]]:S^2\D\times S^2\D\to S^2\D$  is defined on the symmetric product $S^2\D$ as
\be [[A,B]]:=[A',B']\otimes [A'',B''].\ee
Here we use the Sweedler notation $A=A'\otimes A''$, $B=B'\otimes B''$
and we view the self-adjoint operators $\P_\pm$ as the elements of $S^2\D$
in the sense of the formula
\be \P_\pm x:=\P_\pm'(\P''_\pm,x)_\D, \quad x\in \D.\ee

We are now going to show that the affine Poisson T-duality based on the affine Poisson bracket
\eqref{aff} is compatible with the renormalisation group flow if we choose for the Yang-Baxter operators $R^r,R^l$ any pair of the Drinfeld twist operators studied in Section 5.4\footnote{Very recently, there appeared  Ref. \cite{SV} in which it is claimed that the renormalization group flow is compatible with T-duality for every dressing coset. The contents of the present section can be therefore interpreted as an illustration of this general fact.}. In other words,
 we show in the present section, that the subspaces $\E_L,\E_R\in\D_0$ flow in a compatible way. 
 We recall the context: we consider the compact simple Lie algebra  $\K$ equipped with its Killing-Cartan form $(.,.)_\K$ and for its Drinfeld double  we take $\D_\e$ for $\e=0$ (cf. Eq.\eqref{str}). We   pick the Cartan subalgebra $\gH\in\K$ and we consider the subspace $\gH^\perp\subset \K$ which is perpendicular to $\gH$ with respect to the Killing-Cartan form $(.,.)_\K$. 
We recall that any skew-symmetric operator $R:\K\to\K$ is the Drinfeld twist operator, if $\gH^\perp\subset {\rm Ker}(R)$ and Im$(R)\subset\gH$. Any Drinfeld twist operator verifies the Yang-Baxter condition  \eqref{YBl} for $\e=0$  because of the commutativity of the Cartan subalgebra.

Let us solve the flow equation \eqref{flo} with the initial conditions
$\E_L$ and $\E_R$ introduced, respectively, in Eqs. \eqref{ell} and \eqref{err}. We are going to argue that the equation \eqref{flo} implies the following simple flow of the
subspaces 
$\E_L$ and $\E_R$ :
 \be \E_L(s)= \{(a(s)y+R^ry,y),y\in\K\},\label{elt}\ee
 \be \E_R(s)= \{(a(s)y+R^ly,y),x\in\K\},\label{ert}\ee
notably, we remark that only the parameter $a$ flows and its dependence $a(s)$ on the flow parameter $s$ is the same for the case of the initial conditions $\E_L$ as well as  $\E_R$.

Let us first see that the flow $\E_L(s)$   given by Eq.\eqref{elt} fulfils, for a suitable choice of the function $a(s)$, the equation \eqref{flo} with the initial condition $\E_L(0)=\E_L$. For that, pick a basis $t_i$ on the compact simple Lie algebra $\K$ such that 
\be (t_i,t_j)_\K=-\delta_{ij}\label{nr'}\ee
and then  choose the following basis $E_i(s)$ and $E_i^\perp(s)$ of the subspaces $E_{L}(s)$ and $E_L^\perp(s)$ respectively:
\be E_i(s)=\frac{1}{\sqrt{-2a(s)}}\left(a(s)t_i+R^rt_i,t_i\right),  \quad  E_i^\perp(s)=\frac{1}{\sqrt{-2a(s)}}\left(-a(s)t_i+R^rt_i,t_i\right).\label{du2}\ee
Note that it holds
\be (E_i(s),E_j(s))_\D=\delta_{ij}, \quad (E_i^\perp(s),E_j^\perp(s))_\D=-\delta_{ij},  \quad  (E_i(s),E_j^\perp(s))_\D=0\ee
which means that the operators $\E_L(s),\P_\pm(s) :\D\to\D$ viewed as the elements of $S^2\D$ can be written as
\be \E_L(s)=E_i(s)\otimes E_i(s)+E_i^\perp (s)\otimes E_i^\perp (s),\quad \P_+(s)=E_i(s)\otimes E_i(s), \quad  \P_-(s)=-E_i^\perp(s)\otimes E_i^\perp(s)\label{ii}\ee
and the Einstein summation convention holds. 

By differentiating  Eqs. \eqref{du2}, we find
\be \frac{dE_i}{ds}=-\frac{1}{2a}\frac{da}{ds}E_i^\perp, \quad  \frac{dE_i^\perp}{ds}=-\frac{1}{2a}\frac{da}{ds}E_i,\ee
hence
\be \frac{d\E_L(s)}{ds}=-\frac{1}{a}\frac{da}{ds}\left(E_i\otimes E_i^\perp+E_i^\perp\otimes E_i\right)\label{lhs}\ee
Now we use Eqs. \eqref{ii} and calculate the right-hand-side of the flow equation \eqref{flo}
\be \P_+[[\P_+,\P_-]]\P_-+\P_-[[\P_+,\P_-]]\P_+=\left(E_i,[E_k,E^\perp_l]\right)_\D\left(E_j^\perp,[E_k,E^\perp_l]\right)_\D\left(E_j^\perp\otimes E_i+E_i\otimes E_j^\perp\right).\label{bof}\ee
We find  from \eqref{str}, \eqref{bil} and \eqref{du2}
\be \left(E_i,[E_k,E^\perp_l]\right)_\D= -\frac{\sqrt{-2a}}{4}\left(t_i,[t_k,t_l]\right)_\K-\frac{1}{\sqrt{-2a}}\left(t_i,[t_k,R^rt_l]\right)_\K,\ee
\be \left(E_j^\perp,[E_k,E^\perp_l]\right)_\D=-\frac{\sqrt{-2a}}{4}\left(t_j,[t_k,t_l]\right)_\K+\frac{1}{\sqrt{-2a}}\left(t_j,[R^rt_k,t_l]\right)_\K.\ee
We infer from the properties of the Drinfeld twist operators the following identities
\be \left(t_j,[R^rt_k,t_l]\right)_\K \left(t_i,[t_k,R^rt_l]\right)_\K=0,\ee
\be [R^rt_k,[t_k,t_j]]-[t_k,[R^rt_k,t_j]]=0,\ee
which makes possible to rewrite Eq.\eqref{bof} as
\be \P_+[[\P_+,\P_-]]\P_-+\P_-[[\P_+,\P_-]]\P_+=-\frac{a}{8}\left(t_i,[t_k,t_l]\right)_\K \left(t_j,[t_k,t_l]\right)_\K \left(E_j^\perp\otimes E_i+E_i\otimes E_j^\perp\right).\label{bof'}\ee
We have from the normalization condition \eqref{nr} as well as from the ad-invariance of the bilinear form $(.,.)_\K$:
\be \left(t_i,[t_k,t_l]\right)_\K \left(t_j,[t_k,t_l]\right)_\K=-\left([t_i,t_k],[t_j,t_k]\right)_\K=\left(t_i,[t_k,[t_k,t_j]]\right)_\K
=-c\left(t_i,t_j\right)_\K =c\delta_{ij},\ee
where $c$ is the value of the quadratic Casimir in the adjoint representation. The last relation allows us to conclude that
\be \P_+[[\P_+,\P_-]]\P_-+\P_-[[\P_+,\P_-]]\P_+=-\frac{ac}{8}   \left(E_i^\perp\otimes E_i+E_i\otimes E_i^\perp\right).\label{rhs}\ee
Comparing the evaluation of the left-hand-side \eqref{lhs} of the flow equation \eqref{flo} with the evaluation of the right-hand-side \eqref{rhs}, we find the complete agreement provided it holds
\be \frac{1}{a}\frac{da}{ds}=\frac{ac}{8}, \label{fa}\ee
which is the flow equation for  the parameter $a$.  This means that the flowing subspace $\E_L(s)$ given by Eq. \eqref{elt} is the solution of  the
flow equation \eqref{flo} if the function $a(s)$ fulfils the differential equation \eqref{fa}. The remarkable thing is that the exactly analogous calculation with $R^r$ replaced by $R^l$ yields the result that also $\E_R(s)$ given by Eq. \eqref{ert} is the solution of  the
flow equation \eqref{flo} if the function $a(s)$ fulfils the same  differential equation \eqref{fa}. This fact completes the proof of the compatibility of the affine Poisson T-duality with the renormalisation group flow for the case where the Yang-Baxter operators $R^l$ and $R^r$ are the Drinfeld twist operators.

\section{Outlook}
 \setcounter{equation}{0}
It would be interesting to find out  whether there exists
examples of the affine Poisson or of the affine quasi-Poisson T-duality  which would be based on more general Drinfeld doubles than those defined by the commutator \eq{str}. A way to solve this problem would consist in careful inspection of possible affine (quasi)-Poisson structures on an arbitrary  Lie group with the goal to determine whether they admit solutions of the sufficient conditions for the duality formulated in {\bf Assertion 2} of Section 5.2 and in {\bf Assertion 3} of Section 7.2. Another possibility would be to use the dressing coset construction for   more general Drinfeld doubles.

The relation of the affine quasi-Poisson T-duality to the double field theory in the spirit of Ref. \cite{Ha} seems also to be an interesting issue to work out.

\end{document}